\documentclass{aa}  

\usepackage{graphicx}
\usepackage{txfonts}
\usepackage{natbib}
\usepackage{caption}
\usepackage{subcaption}
\usepackage{tablefootnote}

\usepackage{color}
\usepackage[colorlinks=true,linkcolor=blue]{hyperref}

\usepackage{dblfloatfix}
\usepackage{url}
\usepackage{floatrow}
\usepackage{longtable}

\begin{document}

   \title{Constraining the primordial magnetic field with dwarf galaxy simulations}

   \author{Mahsa Sanati
        \inst{\ref{epfl}}, Yves Revaz \inst{\ref{epfl}}, Jennifer Schober \inst{\ref{epfl}}, Kerstin E. Kunze \inst{\ref{Universidad de Salamanca}}, Pascale Jablonka \inst{\ref{epfl}, \ref{Observatoire de Paris}}
          }

   \institute{Institute of Physics, Laboratory of Astrophysics, \'{E}cole Polytechnique F\'{e}d\'{e}ale de Lausanne (EPFL), 1290 Sauverny, Switzerland \label{epfl}\\
        \email{mahsa.sanati@epfl.ch}
    \and Departamento de F\'{i}sica Fundamental, Universidad de Salamanca, Plaza de la Merced s/n, 37008 Salamanca, Spain\label{Universidad de Salamanca}
    \and GEPI, CNRS UMR 8111, Observatoire de Paris, PSL University, 92125 Meudon, Cedex, France \label{Observatoire de Paris}
             }

   \date{Received: XXXX; accepted: YYYY}

\titlerunning{Dwarf galaxies as a probe of Primordial Magnetic Fields}
\authorrunning{M. Sanati et al} 
 
  \abstract
    { 
   Using a set of cosmological hydro-dynamical simulations,
   we constrained the properties of primordial magnetic fields by studying their 
   impact on the formation and evolution of dwarf galaxies.
   We performed a large set of simulations (8 dark matter only and 72 chemo-hydrodynamical) including primordial magnetic fields through the extra density fluctuations they induce at small length scales ($k \geq 10\,h\,\rm{Mpc^{-1}}$) in the matter power spectrum.
   Our sample of dwarfs include 9 systems selected out of the initial $(3.4\,\textrm{Mpc}/h)^3$ parent box, 
   re-simulated from $z = 200$ to $z=0$ using a zoom-in technique and including the physics of baryons.
   We explored a large variety of primordial magnetic fields with strength $B_\lambda$ ranging from $0.05$ to $0.50\,\textrm{nG}$ and magnetic energy spectrum slopes $n_B$ from $-2.9$ to $-2.1$.
   Strong magnetic fields characterized by a high amplitude ($B_\lambda=0.50,\,0.20\,\textrm{nG}$ with $n_B=-2.9$) or by a steep initial power spectrum slope ($n_B=-2.1,-2.4$, with $B_\lambda=0.05\,\textrm{nG}$) induce
   perturbations in the mass scales from  $10^7$ to $10^9\,\rm{M}_{\odot}$. 
   In this context emerging galaxies see their star formation rate strongly boosted. They become more luminous and metal rich than their counterparts without primordial magnetic fields. Such strong fields are ruled out
   by their inability to reproduce the observed scaling relations
   of dwarf galaxies. They predict dwarf galaxies to be at the origin of an unrealistically early reionization of the Universe
   and also overproduce luminous satellites in the Local Group.
   Weaker magnetic fields impacting the primordial density field at corresponding masses 
   $\lesssim 10^6\,\rm{M}_{\odot}$,
   produce a large number of mini dark halos orbiting
   the dwarfs, however out of reach for current lensing observations.
   This study allows for the first time 
   to constrain the properties of primordial magnetic fields based on realistic cosmological simulations of dwarf galaxies.
   }


   \maketitle
%

\section{Introduction}\label{sec:intro}

Magnetic fields have been observed on all cosmic scales probed so far, from planets and stars \citep{2010SSRv..152..651S,Reiners2012} to galaxies \citep{2001SSRv...99..243B, 2013pss5.book..641B} and galaxy clusters \citep{2001ApJ...547L.111C, 2004IJMPD..13.1549G, 2005A&A...434...67V}. There are numerous observational evidences for magnetic fields with a strength of a few to tens of micro Gauss coherent on scales up to ten kpc, detected through radio spectropolarimetry
in spiral galaxies like M51 \citep{2011MNRAS.412.2396F,2015A&ARv..24....4B}, and ultra luminous infrared galaxies (ULIRGs) \citep{2008ApJ...680..981R, 2013ApJ...763....8M}, but also in high redshift galaxies \citep{2008Natur.454..302B, MaoEtAl2017} and in the interstellar and intergalactic medium \citep{2017ARA&A..55..111H}. 
However, understanding the origin and strength of the fields is still a challenge for modern astrophysics. 

There are various theories proposed to explain the generation of magnetic fields on large scales and their amplification by a dynamo in collapsed objects \citep{2006AstHe..99..568I, 2008Sci...320..909R, 2013PhRvL.111e1303N, SchoberEtAl2013}, or the generation of magnetic fields on small scales \citep{2002RvMP...74..775W, 2005ApJ...633..941H, 2018MNRAS.479..315S}. However, recent observational evidences based on blazer emissions suggest that intergalactic medium voids could host a weak $\sim 10^{-16}$ Gauss magnetic field, coherent on Mpc scales \citep{2010Sci...328...73N, 2011PhR...505....1K}.
Such a field is difficult to be purely explained by turbulence in the late universe \citep{2001ApJ...556..619F, 2006MNRAS.370..319B} and would perhaps favor a primordial origin in the early universe \citep{2016RPPh...79g6901S, 2020IAUGA..30..295K, Jedamzik:2020krr}. 

Recently new observational signatures of primordial magnetic fields have been obtained
from the Lyman-$\alpha$ forest clouds \citep{2013ApJ...762...15P}, 
the two-point shear correlation function from gravitational lensing \citep{2012ApJ...748...27P},
the Sunyaev-Zel'dovich statistics \citep{tashiro2009}, 
the cosmic microwave background (CMB) anisotropies \citep{2010PhRvD..81d3517S, 2015A&A...582A..29P, 2017PhRvD..95f3506Z, 2019MNRAS.484..185P}, CMB spectral distortions \citep{Jedamzik:1999bm, Kunze:2013uja, Wagstaff:2015jaa} large scale structure \citep{2010PhRvD..82h3005K} and the reionization history of of the
Universe \citep{pandey2015}.

The above observations provide new upper limits on the strength of the fields which 
appeared to be limited to $47\,\textrm{pG}$ for scale-invariant fields \citep{Jedamzik:2018itu} and a few $\textrm{nG}$ in more general states coherent on $1\, \textrm{Mpc}$. It is thus of great interest to ask if such primordial fields can be confirmed and how their characteristics can be constrained further.

To answer this issue we need to journey back to the early universe. The present day large scale structures are thought to be seeded by quantum field fluctuations when the relevant scales were causally connected, leading to a nearly scale-invariant fluctuation spectrum \citep{dodelson2003modern, 1990S&T....80S.381K, 2002thas.book.....P}. These scales then crossed the universe horizon during the inflationary expansion phase, 
and re-enter only later to serve as the initial conditions, leading to the growth of large scale structures. It is likely that the origin of the magnetic fields also arises during various phase transitions in the early universe \citep{1988PhRvD..37.2743T, 2016JCAP...10..039A, 2019JCAP...10..032D, 2019JCAP...09..008F} by a small fraction of the energy released during the electroweak or quark-hadron transitions and converted to large scale magnetic fields \citep{1983PhLB..133..172H, 1992ApJ...391L...1R}.

A primordial magnetic field (PMF) in the post-recombination epoch generates density fluctuations in addition to the standard inflationary fluctuations. The magnetically-induced perturbations, for a scale-invariant magnetic spectrum, may dominate the standard $\Lambda$CDM matter power spectrum on small length scales and therefore can affect the formation of the first galaxies \citep{1996ApJ...468...28K, 1978PhDT........95W}. 
Indeed, the $\Lambda$CDM paradigm
implies a hierarchical formation of galaxies, in which dwarf galaxies, including the ultra-faint ones (UFDs) \citep{simon2019} as observed today are the best analogs to the smallest initial building blocks.
A modification of the nearly power-law spectrum of the $\Lambda$CDM model at small scales ($k \geq 10\,h \, \rm{Mpc^{-1}}$) will thus directly affect the number and properties of dwarf galaxies with interesting outcomes for understanding the cosmological model and origin of the magnetic field.

Changing the abundance of dwarf galaxies and their structure can potentially shed new light on the long standing tensions existing between dark matter only (DMO) $\Lambda$CDM cosmological
simulations and Local Group observations (see \citet{2017ARA&A..55..343B} for a recent review). Among the existing tensions, the over-prediction of small mass systems around the Milky Way, the so-called 
missing satellite problem \citep{klypin1999,moore1999} and the too big to fail problem \citep{boylankolchin2011,boylankolchin2012} are certainly the most emblematic ones. 
Along with the number of observed satellites, their mass profile is also known to suffer from tensions. DMO simulations predict the dark halos to follow an universal cuspy profile \citep{navarro1996,navarro1997}, while observations favour cored ones (\citet{moore1994,flores1994}, see \citet{read2018} for a short review.)
While improvements in the baryonic treatment of cosmological simulations allowed for a reduction of those tensions \citep{chan2015,sawala2016b,wetzel2016,verbeke2017,revaz2018,pontzen2014,onorbe2015,chan2015,fitts2017, 2019A&A...624A..11H}, modification or alternative to the $\Lambda$CDM have been proposed too,
like warm dark matter \citep{lovell2012,governato2015,fitts2019}, self-interacting dark matter \citep{vogelsberger2014,harvey2018,fitts2019} or wave dark matter \citep{chan2018}.

On the other hand, the plethora of observations available for dwarf galaxies in the Local Group may be used to constrain the amplitude of PMFs.
Any model including the effect of PMFs must reproduce i) 
the abundances of detected satellites around the Milky Way \citep{cautun2018,nadler2018,drlicawagner2019}, ii) the well known observed scaling relations \citep{mateo1998,tolstoy2009,mcconnachie2012,simon2019},
as well as more detailed properties like line-of-sight (LOS) velocity dispersion, stellar abundance patterns and star formation histories
\citep{tolstoy2009}, and iii) they must be in agreement with the Milky Way local reionization history.
Indeed, by hosting the first generation of stars and lightening up the dark ages \citep{2008MNRAS.385L..58C, 2014MNRAS.437L..26S, 2014MNRAS.442.2560W, 2015ApJ...802L..19R, 2015ApJ...811..140B, 2015ApJ...814...69A},
dwarf galaxies played a key role during the epoch of reionization.

The above arguments make dwarf and UFD galaxies excellent laboratories to study the subtle
impact of initial conditions imposed after the inflation and how they rule the first structures.
As such, they may be used to get new constraints on the properties of PMFs \citep{2019ApJ...877L..27S}.

Recently \cite{revaz2018} demonstrated that cosmological simulations run with the code \texttt{GEAR} reproduce a wide range of observed properties of the Local Group dwarf galaxies. 
Relying on this previous work, we take the opportunity
to study for the first time, the formation and evolution of a suite of dwarf galaxies that
emerge from a $\Lambda$CDM cosmology where the primordial density fluctuations include
the effect of PMFs.
Our goal is to size the impact of these primordial fields on the properties of dwarf galaxies.

This paper is organized as follows:
Section~\ref{primordial_magnetic_field}, recalls the theory around PMFs and their effect on the density fluctuation power spectrum. 
Section~\ref{simulations}, describes in details our numerical framework as well as the simulations performed.
The results are presented in Section~\ref{results}.
We first discuss results relying on dark matter only simulations and show
how the PMFs affect the matter halo mass function at small scales. 
In a second part, we focus on zoom-in hydo-dynamical simulations. 
We study the impact of PMFs on the properties of dwarf galaxies
and in particular on scaling relations.
We then estimate how the modified properties of dwarfs directly impact the 
reionization history of the Universe as well as the cumulative number of bright 
satellites around the Milky Way. Limits on the properties of PMFs
are obtained by comparing with existing observational constraints.
A brief conclusion is given in Section~\ref{conclusions}.


\section{Impact of Primordial Magnetic Fields on the density fluctuations}\label{primordial_magnetic_field}
We consider the effect of PMFs generated before recombination by processes occurring during the inflationary epoch 
\citep[ex.][]{turner1988,ratra1992}. 
It has been shown in the seminal works of \citet{wasserman1978} and \citet{1996ApJ...468...28K} that after the recombination epoch,
PMFs may induce motions in the ionized baryons through the Lorentz force which drives compressional and rotational perturbations
causing density fluctuations in the gas. Those perturbations propagate down to other non-ionized components, the neutral gas but also the dark matter
are coupling through the gravitational force 
and a direct impact on the total matter power spectrum is expected .

Characterizing this impact is however non-trivial and a large literature 
has been dedicated to describe it.
Hereafter, without entering into too much detail, we briefly review the principal steps that allow to catch the origin of the effect of primordial 
magnetic fields on the total matter power spectrum.

\subsection{The primordial magnetic field}
Following \citet{wasserman1978,1996ApJ...468...28K,2003JApA...24...51G}, we first assume that the tangled magnetic field results from  a statistically 
homogeneous and isotropic vector random process. The two-point correlation function of a non-helical field in Fourier space can then be expressed as: 
\begin{equation}
    \langle\hat{B}_i(\vec{k})\hat{B}^*_j(\vec{k}')\rangle = (2\pi)^3\delta(\vec{k}-\vec{k}')\frac{P_{ij}(\vec{k},\vec{k}')}{2}P_{B}(k),
    \label{equ:2}
\end{equation}
where $P_{ij}(\vec{k}) = \delta_{ij} - \frac{k_i k_j}{k^2}$,
$k = |\vec{k}|$ is the comoving wave number
and $P_{B}(k)$
is the magnetic field power spectrum.
Without further information on the exact origin of the PMFs, it is usually assumed that its power spectrum follows a simple power law: 
\begin{equation}
  P_{B}(k)=A\,k^{n_B}
  \label{equ:PB}
\end{equation}
where $n_B$ is the slope index, and $A$ the amplitude of the magnetic spectrum, which is defined through the variance of the magnetic field strength $B^2_\lambda$ at a scale $\lambda= 1\,\rm{Mpc} \equiv k_\lambda^{-1}$, namely  \citep{2012PhRvD..86d3510S}:
\begin{equation}
    A = \frac{(2\pi)^{n_B+5}B^2_\lambda}{2\Gamma(\frac{n_B+3}{2})\,k^{n_B+3}_\lambda}.
    \label{equ:A}
\end{equation}{}
With those definitions, $n_B$ and $B_\lambda$ fully characterise the PMFs. They constitute our main free parameters. 
\subsection{Growth of perturbations}\label{perturbation_growth}

In the linearized Newtonian theory, the evolution of density fluctuations in the presence of PMFs is described by the two following coupled equations, 
respectively, for the baryonic fluid perturbation field
$\delta_b(\vec{x},t) = \delta \rho_b(\vec{x},t)/\bar{\rho}_b$ 
and for the collisionless dark matter 
$\delta_{\rm{DM}}(\vec{x},t)= \delta \rho_{\rm{DM}}(\vec{x},t)/\bar{\rho}_{\rm{DM}}$  \citep[see for example][]{1998PhRvL..81.3575S,2005MNRAS.356..778S}:
  \begin{eqnarray}
  \frac{\partial^2}{\partial t^2} \delta_b  + \left[2\,H  + \frac{4\rho_{\gamma}}{3 \rho_b} n_e \sigma_T a   \right] \frac{\partial}{\partial t} \delta_b  & = &  \nonumber \\
  c_b^2 \nabla^2 \delta_b +  4\pi G a^2 \left[ \rho_b \delta_b + \rho_{\rm{DM}} \delta_{\rm{DM}} \right] + \frac{S(\vec{x}, t)}{a^3},
  \label{equ:1a}
  \end{eqnarray}
and
  \begin{equation}
  \frac{\partial^2}{\partial t^2} \delta_{\rm{DM}} + 2\,H   \frac{\partial}{\partial t} \delta_{\rm{DM}}  =
  4\pi G a^2 \left[ \rho_b \delta_b + \rho_{\rm{DM}} \delta_{\rm{DM}} \right].
  \label{equ:1b}
  \end{equation}
%
\begin{figure*}[t]
    \centering
    \begin{floatrow}{}
    \ffigbox
    {\includegraphics[width=.5\textwidth]{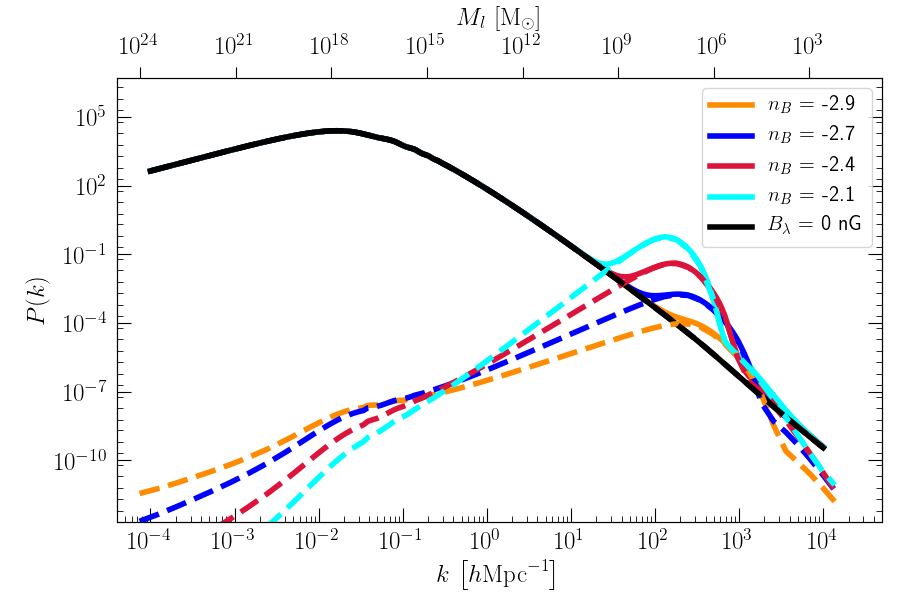}}
    {\caption{The contribution of the magnetically induced power spectrum for a variety of magnetic field indexes (dashed colored lines) to the total matter power spectrum (plain colored lines). The unperturbed $\Lambda$CDM spectrum is show in the black continuous line. The magnetic field strength is kept constant at $B_\lambda = 0.05\,\textrm{nG}$.} 
    \label{fig:powerspectrumBl}}
    \ffigbox
    {\includegraphics[width=.5\textwidth]{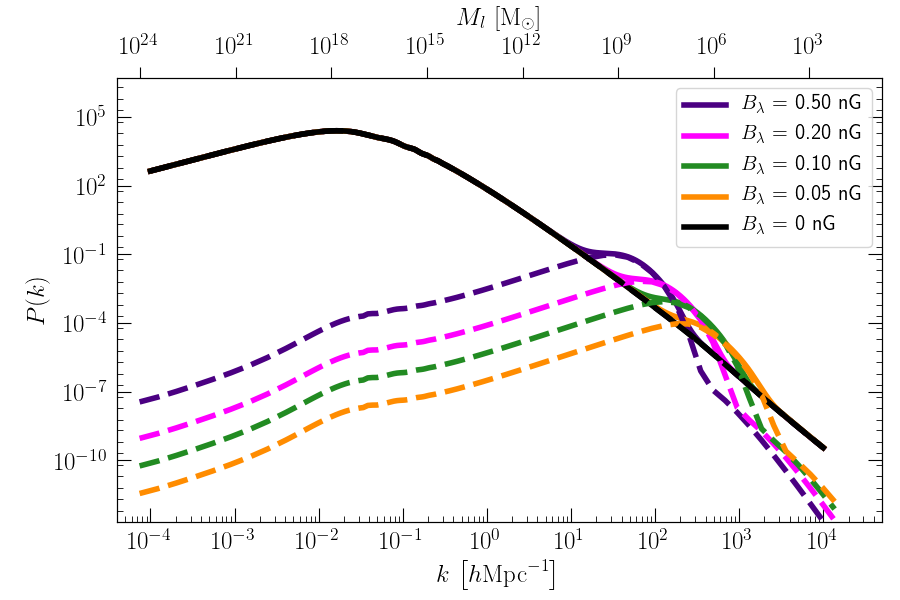}}
    {\caption{The contribution of the magnetically induced power spectrum  to the total matter power spectrum. Same figure as Fig.~\ref{fig:powerspectrumBl} with constant spectral slope, $n_B = -2.9$, and varying magnetic field amplitudes.}
    \label{fig:powerspectrumnB}}
    \end{floatrow}
\end{figure*}

Here, $a$ is the scale factor, $H = \dot{a}/a$ the Hubble constant, 
$\rho_b$ and $\rho_\gamma$ the baryon and photon mass density, $n_e$ the electron number density, $\sigma_T$ the Thomson cross-section for electron-photon scattering,
and $c_b$ the baryon sound speed. 
$S$ represents the magnetic field source term  normalized to the baryon density at the present time, $\overline{\rho}_b(t_0)$:
\begin{equation}
S(\vec{x}, t) = \frac{\pmb{\nabla}\cdot[\boldsymbol{B}\times(\pmb{\nabla}\times\boldsymbol{B})]}{4\pi\overline{\rho}_b(t_0)},
\label{equ:2}
\end{equation} 

The baryon pressure term $c_b^2 \nabla^2 \delta_b$ in Eq.~\ref{equ:1a} is sub-dominant with respect to the magnetic pressure as long as the background magnetic field
is larger than $5\times 10^{-11}~\rm{G}$ \citep{1998PhRvL..81.3575S} and may thus be ignored.
The damping term involving the Thomson cross-section corresponds to the radiation viscosity. Prior to recombination,
this term leads to the damping of small scale magnetic waves \citep{jedamzik1998,1998PhRvL..81.3575S}, a physical process similar to the Silk damping \citep{1968ApJ...151..459S}.
This induces a sharp cutoff of the magnetic field when entering the recombination epoch and subsequently, to its contribution in the total matter power spectrum.

Introducing the total matter density perturbation:
\begin{equation}
\delta_m(\vec{x},t)=(\overline{\rho}_{\rm{DM}}\delta_{\rm{DM}}+\overline{\rho}_b \delta_b)/\overline{\rho}_{\rm{DM}},
\label{equ:delta_m}
\end{equation}
we can solve Eqs.~\ref{equ:1a} and \ref{equ:1b}  and get the time evolution of $\delta_m$ 
only due to the magnetic fields \citep{2005MNRAS.356..778S}:
\begin{equation}
\delta_m \sim \frac{3}{5}\frac{\Omega_b}{\Omega^2_m}\left[  \frac{3}{2}\left(   \frac{t}{t_{\rm{rec}}}  \right)^{2/3} + \left( \frac{t_{\rm{rec}}}{t}\right) - \frac{5}{2}     \right] S(\vec{x},t_{\rm{rec}})\, t^2_{\rm{rec}},
\end{equation}
where $t_{\rm{rec}}$ is the time at recombination and $\Omega_m$ and $\Omega_b$ are the matter and baryon density parameters.

The important point to get from the previous equations is that the spatial dependence of $\delta_m(\vec{x},t)$ can be followed through the magnetic field source term 
$S(\vec{x},t_{\rm{rec}})$. We thus expect the total matter power spectrum to directly depend on the power spectrum of the magnetic field.

It is worth noting that the most significant evolution of the magnetic field
takes place before recombination \citep{KahniashviliEtAl2013,BrandenburgEtAl2017,BrandenburgEtAl2017b}. During recombination the ionization degree drops to a tiny value. Afterwards the magnetic field is more or less frozen into the gas, i.e., it just follows passively the cosmic expansion, ensuring the magnetic flux to be conserved.

\subsection{Impact on the total matter power spectrum}

The modification of the total matter power spectrum due to the influence of magnetic fields has been first addressed by \citet{1996ApJ...468...28K} and extended by 
\citet{2003JApA...24...51G}.
The total matter power spectrum is the ensemble average of the density fluctuations in the Fourier space, $P(k,t) = \langle \delta_m(k,t)\,\delta^*_m(k,t) \rangle$, that we can obtain from the evolution equation (Eq.~\ref{equ:delta_m}) and introducing the ensemble average of the PMFs, Eq.~\ref{equ:2} and \ref{equ:PB}.

After the decoupling of photons, the ionized matter density fluctuations are affected by the magnetic Alfv\'en waves if the crossing time, $\tau_A \sim 1/{k\,v_A}$, is smaller than the inverse Hubble rate \citep{2004PhRvD..70l3003B}, where $v_A$ is the Alfv\'en velocity 
$v_A = {B_\lambda}/{\sqrt{\mu_0\rho_0}}$, with $\mu_0$ the  permeability or magnetic constant.
As the magnetic energy $B_{\lambda}^2$ scales with $P_B(k)\,k^3$, $\tau_A \sim \sqrt{\mu_0 \rho_0}\,k^{-\left( 3+n_B \right)/2}$. For $n_B>-5$, the crossing time is then shorter for larger $k$, 
amplifying the perturbations faster at smaller scales.
In the limit where $k \ll k_B$, with $k_B \cong \left( v_{\rm{A}} \sqrt{\pi/\rho_0 G} \right)^{-1}$, the magnetic wave number above it the Alfv\'en waves damp instabilities 
(the equivalent of the Jeans wave number), at lowest order in  $k/k_{\rm{max}}$, i.e., $k_{\rm{max}} \cong  k_B$,  the solution for the total matter power spectrum is \citep{2003JApA...24...51G}:
\begin{equation}
  P(k) \sim A k^{2 n_B + 7} + B k_{\rm{max}}^{2 n_B + 3} k^4 + C k_{\rm{max}}^{2 n_B + 1} k^6 + ...
\label{equ:pk}  
\end{equation}
which strongly depends on the magnetic slope index $n_B$. For the scale-invariant case, $n_B \cong -3$, $P(k) \sim k$, and $\tau_A \sim k^{-1}$, thus small scale perturbations are strongly
amplified, as long as $\tau_A < 1/H(t)$, up to about $k_B$ where they are sharply quenched (see \ref{perturbation_growth}). 

\begin{figure*}[t]
    \centering
    \includegraphics[width=\textwidth]{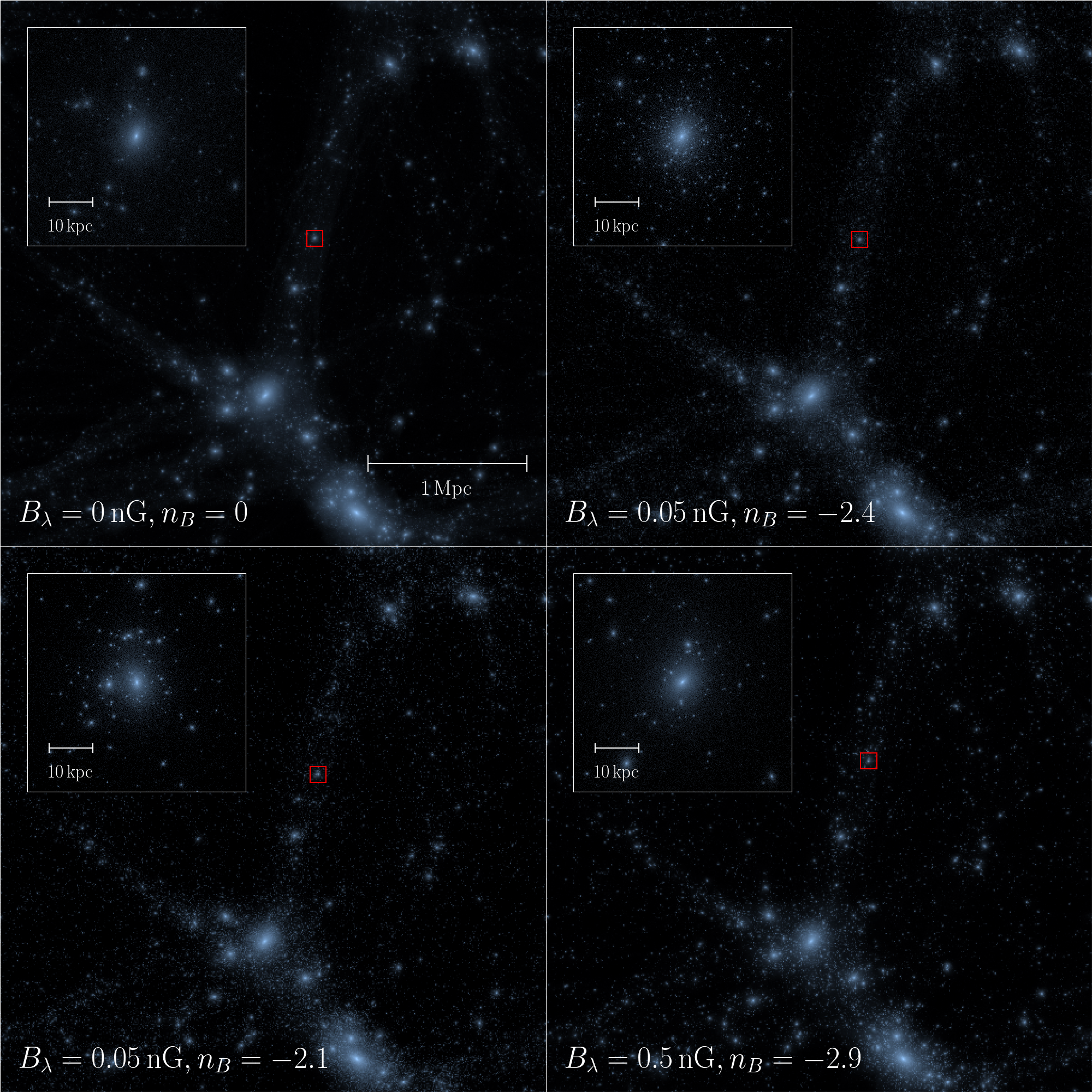}
    \caption{Surface density projections of three DMO simulations at $z=0$ with different PMFs as indicated on the bottom. In all plots, the upper square is a zoom on a dwarf indicated by the red central squares.}
    \label{fig:DMO}
\end{figure*}
\subsection{The adopted total matter power spectrum}

We generated the matter power spectrum using a modified version of the \texttt{CAMB} code which includes
the effects of PMFs (for details see \cite{2012PhRvD..86d3510S}). In this version, the non-linear effects of both 
the magnetic Jeans length (magnetic pressure) and the damping due to the radiation viscosity (when the photon free streaming length is small) is explicitly computed.
As such, contrary to other studies, there is no need to artificially include a cutoff wave number.

Figures~\ref{fig:powerspectrumBl} and \ref{fig:powerspectrumnB} display the matter power spectrum due to the magnetic field (dashed lines)
together with its contribution to the total matter power spectrum (plain lines). The black continuous line corresponds to the
unperturbed $\Lambda$CDM power spectrum.
The dependence with respect to the magnetic field amplitude ($B_\lambda = 0.05,\ 0.10,\ 0.20,\ 0.50\,\textrm{nG}$) is shown in Figure~\ref{fig:powerspectrumBl} keeping the spectral index constant 
($n_B\ =\ -2.9$). The dependence with respect to the spectral index with $B_\lambda = 0.05\,\rm{nG}$ is shown on Fig.~\ref{fig:powerspectrumnB}.
The range of parameters explored is chosen in such a way that they impact the power spectrum without violating
the existing constraints as the ones presented in Section~\ref{sec:intro}.  

It is worth mentioning that, for the magnetic spectral index, a range of both positive and negative values are suggested. In this study, we considered PMFs generated during the inflation, which tend to have negative values for $n_B$, \citep{1988PhRvD..37.2743T, 1992ApJ...391L...1R, 2000PhRvD..62j3512G, 2008JCAP...01..025M}.
However, positive values for $n_B$ are also plausible in other magnetogenesis processes, like the electroweak phase transition \citep{2001PhR...348..163G, 2011PhR...505....1K, 2013A&ARv..21...62D}, Higgs field gradients \citep{1991PhLB..265..258V}, or in the case of subsquent evolution of the PMFs, e.g.\ via the chiral anomaly \citep{2012PhRvL.108c1301B, RogachevskiiEtAl2017, SchoberEtAl2018}.

As expected from Eq.~\ref{equ:pk}, at large scales (small $k$), for $n_B = -2.9$ the slope is nearly 1 and increases for increasing $n_B$.
Small scale structures are thus strongly magnetically amplified and reach an amplitude larger than the one induced by inflation only. At very small scales, 
the power spectrum is sharply quenched owing to the magnetic pressure and radiation viscosity damping. 
The total power spectrum is thus characterised by a bump at the smallest amplified scales, $k \sim 10 - 1000\,h\rm{Mpc^{-1}}$ depending on the exact values of
$n_B$ and $B_\lambda$.
In Fig.~\ref{fig:powerspectrumBl} and \ref{fig:powerspectrumnB} we link the wave number and the mass contained within a sphere of comoving Lagrangian radius $r_l$ at $z = 0$ by defining the mass scale $M_l$ (the equivalent of the Jeans mass), 
assuming a background mean density given by the cosmological parameters \citep[see for example][]{2017ARA&A..55..343B}: 
\begin{equation}
M_l = \frac{4\pi}{3}r^3_l \rho_m = \frac{\Omega_m H^2_0}{2G}r^3_l = 1.71\times10^{11}\rm{M}_{\odot}(\frac{\Omega_m}{0.3})(\frac{h}{0.67})^2(\frac{r_l}{1\textrm{Mpc}})^3    
\end{equation}
The bumps precisely cover masses expected for the total mass of
dwarf galaxies. Therefore, PMFs could significantly influence the formation and properties of those objects.

From Fig.~\ref{fig:powerspectrumBl} and \ref{fig:powerspectrumnB},
it appears that a steeper slope or a stronger amplitude would
affect the power spectrum for $k<10\,h \rm{Mpc^{-1}}$. Such a modification
is ruled out by the observations of the Lyman-$\alpha$ forest \citep{tegmark2002,2013ApJ...762...15P}.

\section{Simulations}\label{simulations}
\begin{table*}[t]
    \centering
    \begin{tabular}{c  c c c c  c c c c}
    \hline\hline
    model & \texttt{B0.05n2.9} & \texttt{B0.05n2.7} & \texttt{B0.05n2.4} & \texttt{B0.05n2.1} &  \texttt{B0.10n2.9} & \texttt{B0.20n2.9} & \texttt{B0.50n2.9}\\
    \hline
    $B_\lambda\,\rm{[nG]}$  & 0.05 & 0.05 & 0.05 & 0.05 &  0.10 & 0.20 & 0.50\\ 
    $n_B$ & -2.9  & -2.7  & -2.4  & -2.1  & -2.9 & -2.9 & -2.9 \\
    \hline
    \end{tabular}
    \caption{The combination of explored parameters $B_\lambda$ and $n_B$ used for the seven simulations performed with a magnetically perturbed power spectrum. The models are referred in the paper as the model names given below.}
    \label{tab:parameters}    
\end{table*}    
We performed a set of $\Lambda$CDM cosmological simulations with initial conditions 
that match the perturbed range of the total matter power spectrum described above.
We used the same $(3.4\,\rm{Mpc}/h)^3$ box adopted by \citet{revaz2018} to study dwarf 
galaxies. The cosmology is the one described by the \cite{2016A&A...594A..13P} with: $\Omega_\Lambda = 0.685,\ \Omega_m = 0.315,\ \Omega_b = 0.0486,\ H_0 = 67.3\ \textrm{km s}^{-1} \textrm{Mpc}^{-1},\ n_s = 0.9603,\ \sigma_8 = 0.829$.

Two types of simulations have been realized: (i) Dark Matter Only (DMO) simulations of the 
entire box, used to compute the halo mass function and its dependency on the PMFs (ii) hydro-dynamical simulations of a selection of dwarf galaxies with the zoom-in technique. 
They allow to check the effect of PMFs on the properties of stellar populations.

\subsection{Initial conditions}

The initial conditions have been generated using the code \texttt{MUSIC} \citep{2011MNRAS.415.2101H}.
Instead of using, the \cite{1998ApJ...496..605E}, the default \texttt{MUSIC} power spectrum, we designed a special plug-in to use the total matter power spectrum obtained from the modified \texttt{CAMB} code. 
Compared to the Eisenstein \& Hu function, the latter includes baryonic pressure which leads to 
additional power at scales larger than $k \cong 100\,h\,\rm{Mpc}^{-1}$, and results in almost a 10 \% difference in the mass content of structures formed in models without a magnetic fields.

For the DMO simulations we used the resolution of level 9. This corresponds to $(2^9)^3$ particles, covering the entire box. For the hydro-dynamical simulations we used the zoom-in technique to gradually degrade the resolution from level 9 to level 6 outside the refined regions.
With this setting, the mass resolution of dark matter, gas and stellar particles is $22'462\,h^{-1}\rm{M}_{\odot}$, $4'096\,h^{-1}\rm{M}_{\odot}$ and $1'024\,h^{-1}\rm{M}_{\odot}$, respectively.

\begin{table*}[b]
    \centering
    \begin{tabular}{c  c c c c  c c c c}
    \hline\hline
    model & \texttt{B0.05n2.9} & \texttt{B0.05n2.7} & \texttt{B0.05n2.4} & \texttt{B0.05n2.1} &  \texttt{B0.10n2.9} & \texttt{B0.20n2.9} & \texttt{B0.50n2.9}\\
    \hline
    c & 3.00  & 6.71  & 7.79 & 7.35  & 4.50 & 4.06 & 3.86 \\
    u & 5.30  & 6.14  & 7.07  & 7.85  & 6.17 & 7.08 & 8.45 \\
    s & 0.80  & 0.60  & 0.57  & 0.52  & 0.70 & 0.70 & 0.67 \\
    \hline
    \end{tabular}
    \caption{Values of the fitting parameters used in Eq.~\ref{eq:dNdMdNdM} for different PMF models.}
    \label{tab:cus}    
\end{table*}    

\subsection{Code evolution}

The simulations have been run with the chemo-dynamical Tree/SPH code \texttt{GEAR} developed by \cite{2012A&A...538A..82R}, \cite{2016A&A...588A..21R} and \cite{revaz2018}.
\texttt{GEAR} is a fully parallel code based on \texttt{Gadget-2} \citep{2005MNRAS.364.1105S}. It operates with individual and adaptive time steps \citep{2012MNRAS.419..465D} and includes recent SPH improvements such as the pressure-entropy formulation \citep{2013ascl.soft05006H}.
\texttt{GEAR} includes radiative gas cooling and redshift-dependent UV-background heating through the \texttt{GRACKLE} library \citep{2017MNRAS.466.2217S}. The metal line cooling is computed through the Cloudy code \citep{2013RMxAA..49..137F} for solar abundances which is scaled according to the gas metallicity. Hydrogen self-shielding against the ionizing radiation is incorporated by suppressing the UV-background  heating for gas densities above $n_H = 0.007$ $\mathrm{cm}^{-3}$ \citep{2010ApJ...724..244A}.
Star formation relies on the stochastic prescription proposed
by \citet{katz1992,katz1996}. We used an efficiency of star formation parameter $c_\star = 0.01$.
The star formation recipe is supplemented by a modified version of the Jeans pressure floor 
\citep{2011MNRAS.417..950H,revaz2018} 
by adding a non-thermal term in the equation of state of the gas to avoid any spurious gas fragmentation. 
The chemical evolution includes Type Ia and II supernovae (SNe) with yields taken from \cite{1995MNRAS.277..945T} and \cite{2000ApJ...539...26K}, respectively.
\texttt{GEAR} includes thermal blastwave-like feedback, for which 10\% of the explosion energy of each SN, taken as $10^{51} \textrm{erg}$, is released in the interstellar medium (i.e., the SN efficiency is $\epsilon = 0.1$).
The initial mass function (IMF) is sampled with the random discrete scheme (RIMFS) of \cite{2016A&A...588A..21R}. The released chemical elements are mixed using the Smooth Metallicity Scheme 
\citep{okamoto2005,tornatore2007,wiersma2009}.

\subsection{The set of simulations}
Each type of simulations, DMO and zoom-in, have been run eight times. 
A first run is performed without magnetically-induced perturbations and
seven others explore the effect of $B_\lambda$ and $n_B$ parameters with values given in
Tab.~\ref{tab:parameters} and corresponding to the total matter power spectra of Fig.~\ref{fig:powerspectrumBl} and \ref{fig:powerspectrumnB}. 

We selected nine halos from the 27 dwarfs presented in \cite{revaz2018} and re-simulated them with the same zoom-in technique.
The selection covers galaxies with $L_{\rm{V}} < 10^7\,\rm{L}_{\odot}$, where $L_{\rm{V}}$ stands for the total V-band luminosity, and different star formation histories. 
Seven dwarfs are dominated by an old stellar population, 
being quenched by the UV-background after at most four billion
years and two dwarfs are more massive, characterized by extended star formation histories.
The list of the re-simulated dwarf galaxies is given in Tab.~\ref{tab:halos} with their basic properties 
as obtained in the unperturbed case at $z=0$: total V-band luminosity $L_{\rm{V}}$, total stellar mass $M_{\star}$
and virial mass $M_{200}$, where the virial overdensity is 200 times the critical density.

All simulations have been started at a redshift of 200, ensuring that the rms variance of the initial density field, $\sigma_8$, lies between 0.1 and 0.2 \citep{2009ApJ...698..266K}. At the exception of the extreme cases, all halos
reached $z=0$ (the reason of the crash will be discussed later on). 
\begin{table}[h]
    \begin{center}
    \begin{tabular}{c c c c c}
      \hline
      \hline 
      model & $L_{\rm{V}}$ & $M_{\star}$ & $M_{200}$ &  SF Class \\ 
      ID    & [$10^6 \rm{L}_{\odot}$] & [$10^6 \rm{M}_{\odot}$] & [$10^9 \rm{M}_{\odot}$]  \\ 
      \hline 
      \texttt{h050}     & 4.2 & 9.6 & 2.6 &  Extended \\
      \texttt{h070}     & 2.0 & 5.8 & 1.8 &  Extended \\
      \texttt{h061}     & 0.2 & 0.5 & 1.9 &  Quenched \\
      \texttt{h141}     & 0.2 & 0.6 & 0.8 &  Quenched \\
      \texttt{h111}     & 0.2 & 0.5 & 1.1 &  Quenched \\
      \texttt{h122}     & 0.1 & 0.4 & 1.0 &  Quenched \\
      \texttt{h159}     & 0.4 & 1.1 & 0.7 &  Quenched \\
      \texttt{h168}     & 0.1 & 0.3 & 0.6 &  Quenched \\
      \texttt{h177}     & 0.2 & 0.5 & 0.5 &  Quenched \\
      \hline
    \end{tabular}
    \end{center}
    \caption{The list of the 9 dwarf galaxies simulated with the zoom-in technique. For each of them, we provide the total V-band luminosity $L_{\rm{V}}$,
    total stellar mass $M_{\star}$ and virial mass $M_{200}$ and the star formation type, for the unperturbed models at $z=0$.}
    \label{tab:halos}    
\end{table}

\section{Results}\label{results}
\begin{figure*}[t]
    \centering
    \begin{floatrow}
    \ffigbox
    {\includegraphics[width=.5\textwidth]{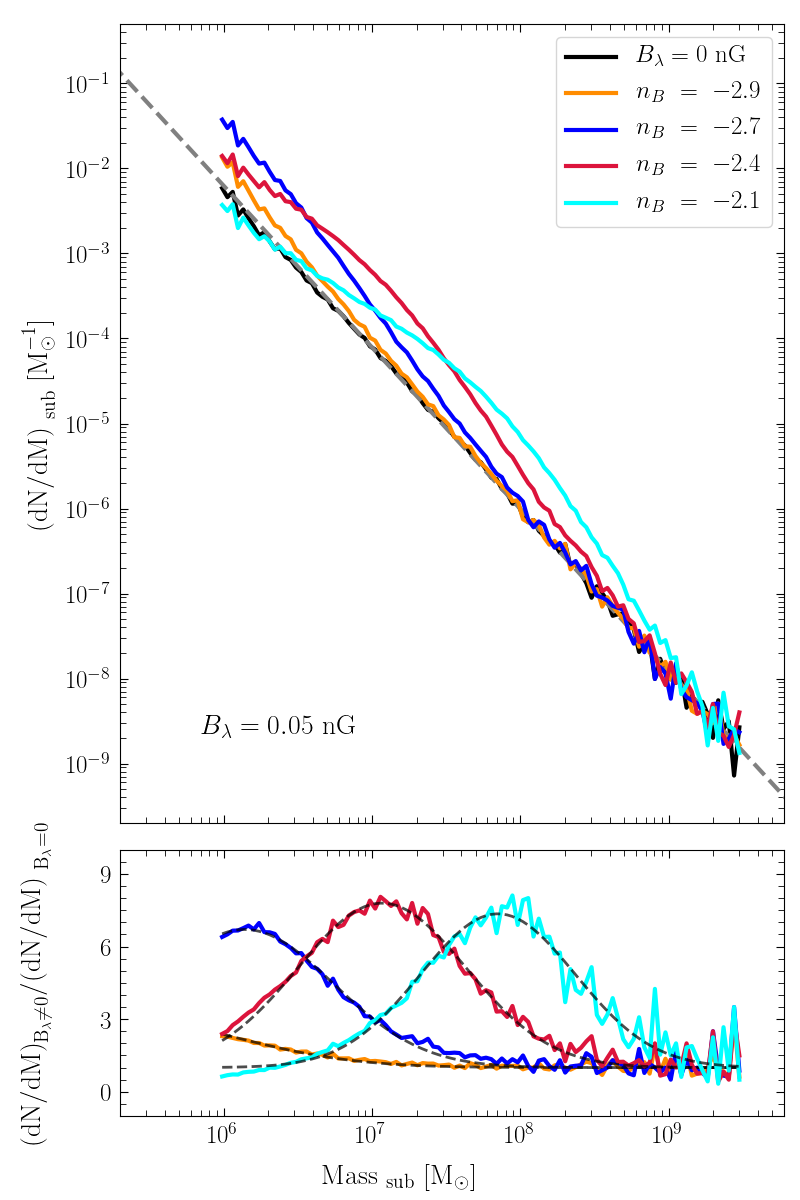}}
    {\caption{Halo mass function. The halo mass is calculated within $R_{200}$. We show the count of subhalos per logarithmic mass interval for different slope indexes $n_B$ with a constant magnetic field strength, $B_\lambda = 0.05 \textrm{nG}$. 
    The dashed gray line shows a power-law $dN/dM \sim M^{-1.96}$ fitted the unperturbed power spectrum. The bottom panel shows the ratio between the halo mass function of each PMF model and the unperturbed model. The dashed lines correspond to the fit given by Eq.~\ref{eq:dNdMdNdM}.}
    \label{fig:dNdMa}}
    \ffigbox
    {\includegraphics[width=.5\textwidth]{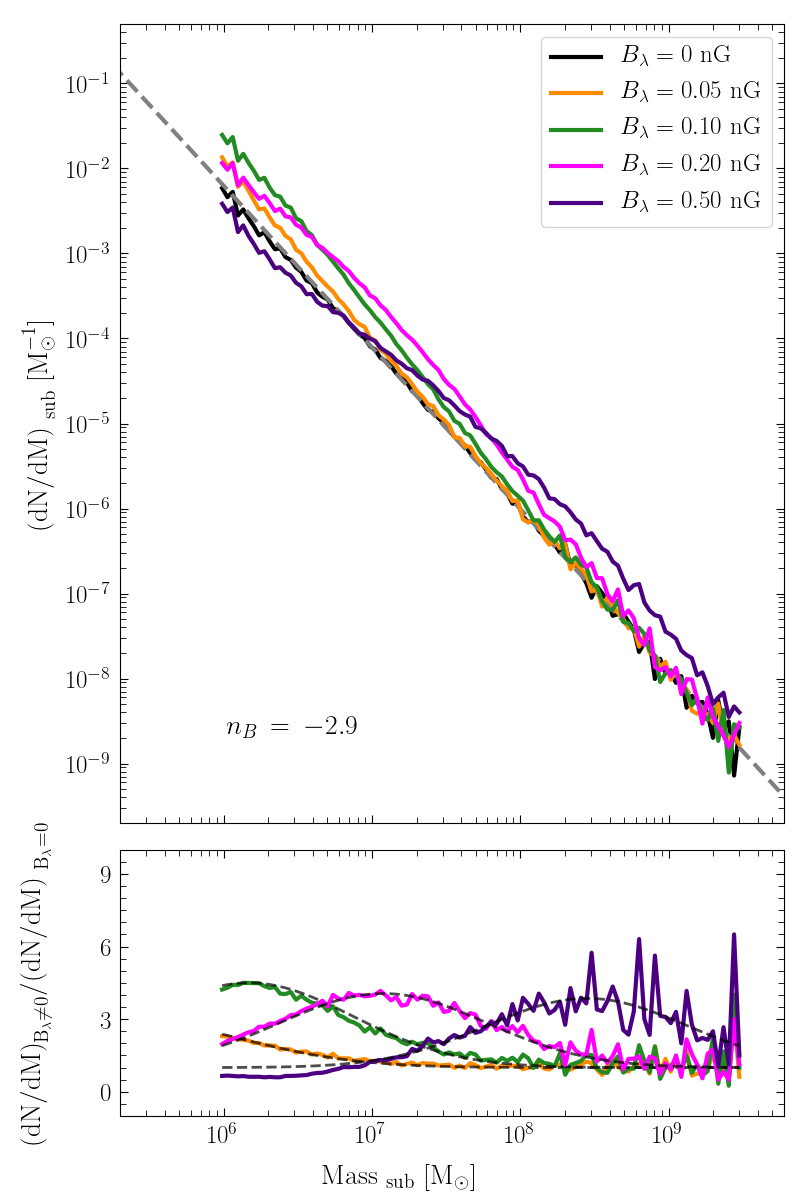}}
    {\caption{Halo mass function. Same figure as Fig.~\ref{fig:dNdMa} for varying magnetic field strengths with the spectral slope kept constant to $n_B = -2.9$.}
    \label{fig:dNdMb}}
    \end{floatrow}
\end{figure*}

\subsection{DMO simulations and halo mass function}\label{sec:halo_mass_fct}

Figure~\ref{fig:DMO} displays the dark matter surface density at $z=0$ of four models:
the unperturbed $\Lambda$CDM model and three models with the magnetically induced bumps in the power spectrum that peak at different mass scales:
\texttt{B0.05n2.4} ($M_{\rm{peak}}\cong 5\cdot 10^6\,\rm{M}_{\odot}$), \texttt{B0.05n2.1} ($M_{\rm{peak}}\cong 3\cdot 10^7\,\rm{M}_{\odot}$), and \texttt{B0.50n2.9} ($M_{\rm{peak}}\cong 2\cdot 10^8\,\rm{M}_{\odot}$).
As expected, the small scale bump increases the number of sub-dark halos orbiting
around dwarf galaxies. This number decreases with increasing the mass of the halos.
To quantify the effect of the bump observed in the total matter power spectrum, for all our DMO simulations, we investigate the abundance of dark matter sub-halos. We extracted dark halos using the
\texttt{Rockstar} halo finder \citep{2013ApJ...762..109B} that uses  an adaptive hierarchical refinement of the friends-of-friends algorithm.
All extracted halos consist of particle groups that are over-dense with respect to the local background and contain a minimum of 100 bounded particles.

Fig.~\ref{fig:dNdMa} and \ref{fig:dNdMb} show the halo mass function, i.e., the number of dark matter halos, $dN$, per unit mass interval, $dM$. All mass functions are truncated below
$3\cdot 10^5\,\rm{M}_{\odot}$, which corresponds to the limit of our mass resolution.
While halos are detected with masses up to about $5\cdot 10^{10}\,\rm{M}_{\odot}$, 
the relatively small size of the box prevents to generate many of those. 
This makes the curve noisy and we truncated it
above $2\cdot 10^9\,\rm{M}_{\odot}$. Between these limits, the unperturbed power spectrum is nicely 
fitted by a power-law, 
\begin{equation}
    \frac{dN}{dM} = a_0\,\left( \frac{M}{m_0} \right)^\alpha,
    \label{eq:dNdM}
\end{equation}
with an amplitude of $a_0 = 8.79 \times 10^7 / M_{vir} = 4.46 \times 10^{-5}{\rm{M}_{\odot}}^{-1}$, 
a pivot point of $m_0 = 10^{-5} M_{vir} = 1.97 \times 10^7\, \rm{M}_{\odot}$ and
a slope of $\alpha = -1.96$. Those parameters perfectly match previous studies \citep[see for example][]{springel2008}.
The fit to the halo mass function is shown in dashed gray line.

The bottom panels of Fig.~\ref{fig:dNdMa} and \ref{fig:dNdMb} show the ratio between the magnetically modified halo mass function and the unperturbed one.
Depending on the strength and the spectral index of the magnetic field, each PMF model affects the halo mass function  in a different mass range 
below $\sim 10^9\, \rm{M}_{\odot}$, reflecting the corresponding mass of the bump in the power spectrum.
For a fixed magnetic amplitude, increasing the slope from $n_B\,=\,-2.7$ to $-2.1$, shifts the bump from about $10^6\,\rm{M}_{\odot}$
to $5\cdot 10^7\,\rm{M}_{\odot}$. In average, in a one dex mass range around the maximum of the bump, between 5 to 7 times more halos are present compared to the generic model without magnetic fields.
Similarly, when the slope is fixed, increasing the magnetic amplitude, $B_\lambda$, shifts the bump upwards, increasing their number by a factor between 3 and 4.

We found the ratio between the perturbed and unperturbed halo mass functions to be well fitted
by a simple modified Gaussian function:
\begin{equation}
    \left(\frac{dN}{dM}\right)/ \left (\frac{dN}{dM}\right)_{\rm{unperturbed}}  =  (c-1)\, \exp\left[{-\frac{\left(\log_{10}(M)-u\right)^2}{2\,s^2}}\right]  + 1
    \label{eq:dNdMdNdM}
\end{equation}
with values for $c$, $u$ and $s$ given in Tab.~\ref{tab:cus}. The fit is shown by a dashed line on the bottom panel of Fig.~\ref{fig:dNdMa} and \ref{fig:dNdMb}.
Equation~\ref{eq:dNdMdNdM} is convenient to estimate analytically the magnetically perturbed halo mass
function from a known unperturbed one. This will be used in Sec.~\ref{sec:lg_satellites} to compare the predicted number of Milky Way satellites to the observed ones.

\subsection{Zoom-in simulations: physical properties}
\begin{figure*}[th!]
    \centering
    \includegraphics[width = .99\textwidth]{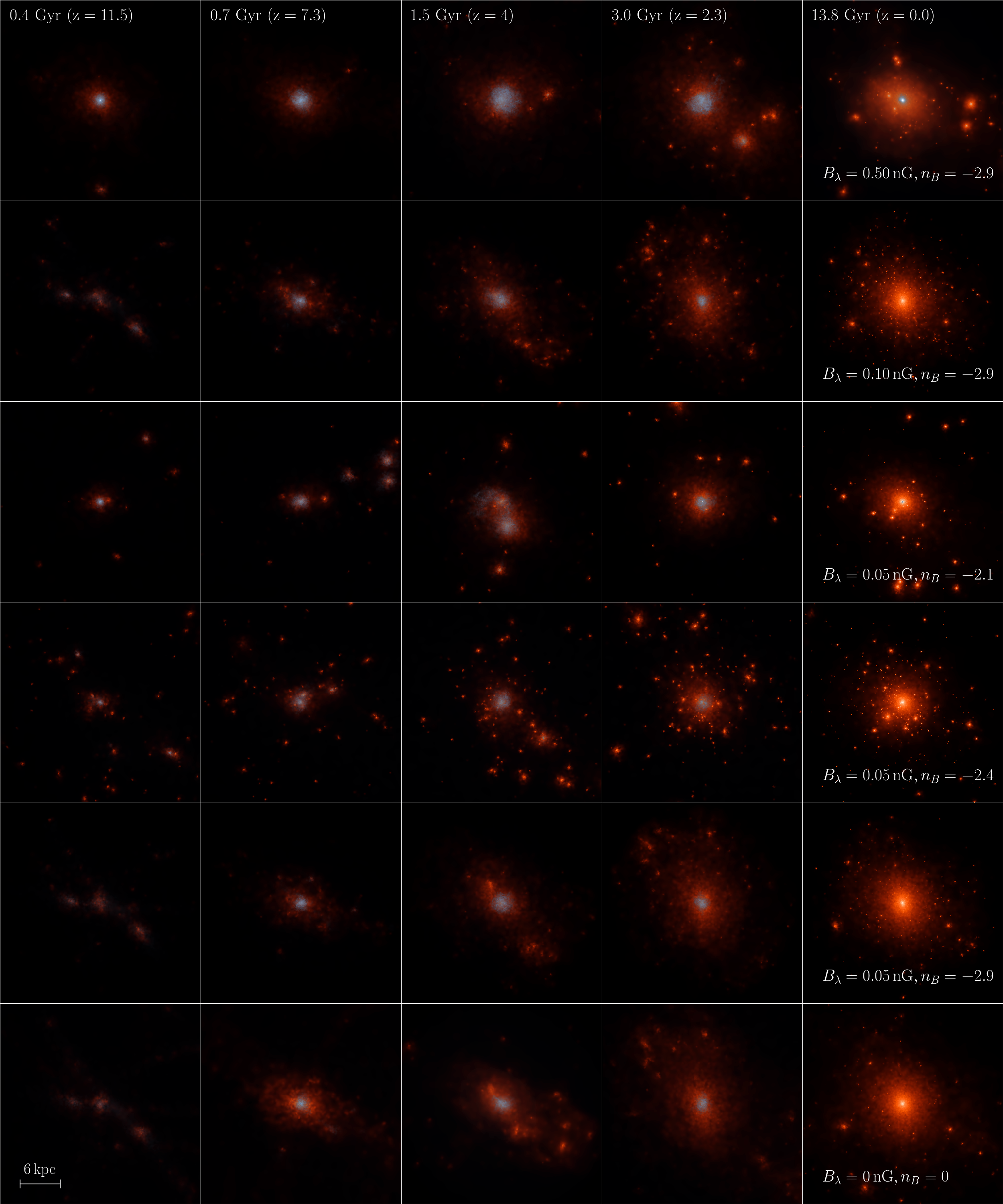}
    \caption{Time evolution of halo \texttt{h070} from $z=11$ to $z=0$ for different magnetically perturbed  models. The unperturbed model is shown at the bottom for comparison.
    In all images stars are shown in white, the gas is displayed in blue, and the dark matter in orange.
    Each box is 20 comoving kpc on a side.}
    \label{fig:time_evolov}
\end{figure*}
\begin{figure*}[t]
    \centering
    \begin{floatrow}
    \ffigbox
    {\includegraphics[width=.5\textwidth]{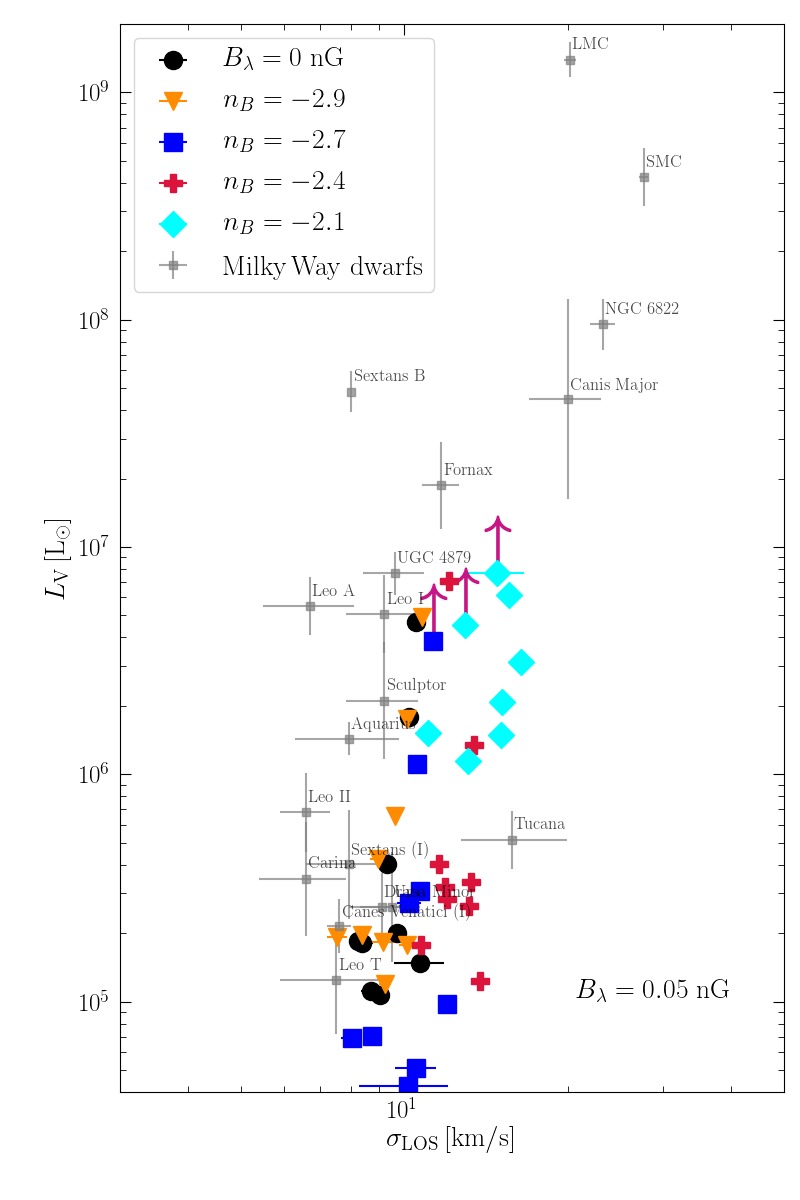}}
    {\caption{V-band Luminosity as a function of line-of-sight velocity dispersion in each model galaxy compared to the observational data of Milky Way satellites in black squares (see Sect.~\ref{scaling_relations} for references).
    The purple arrows indicate models that did not reach $z=0$ and are expected to have a slightly brighter luminosity.}
    \label{fig:LvvsSigmaBl}}
    \ffigbox
    {\includegraphics[width=.5\textwidth]{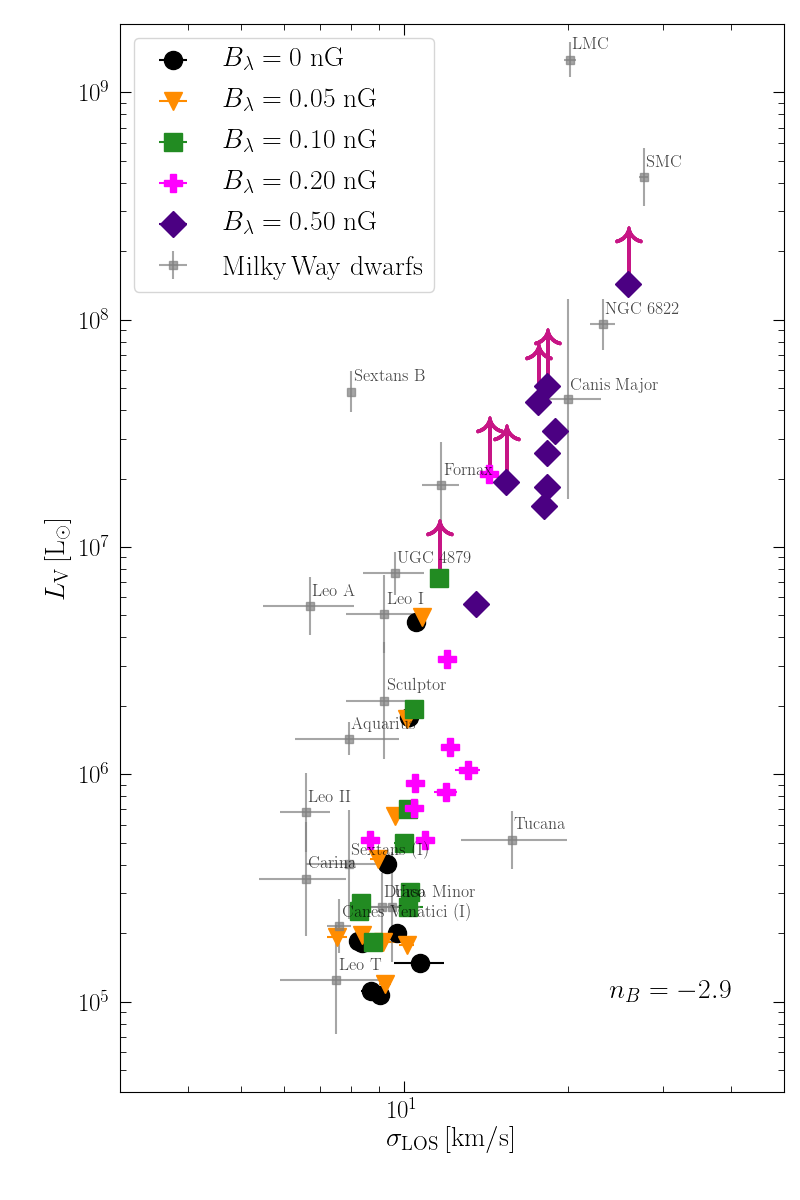}}
    {\caption{V-band Luminosity as a function of line-of-sight velocity dispersion. Same figure as Fig.~\ref{fig:LvvsSigmaBl} but for models with constant spectral slope and varying
    magnetic field strengths $B_\lambda$.}
    \label{fig:LvvsSigmanB}}
    \end{floatrow}
\end{figure*}{}


\begin{figure*}[t]
    \centering
    \begin{floatrow}
    \ffigbox
    {\includegraphics[width=.5\textwidth]{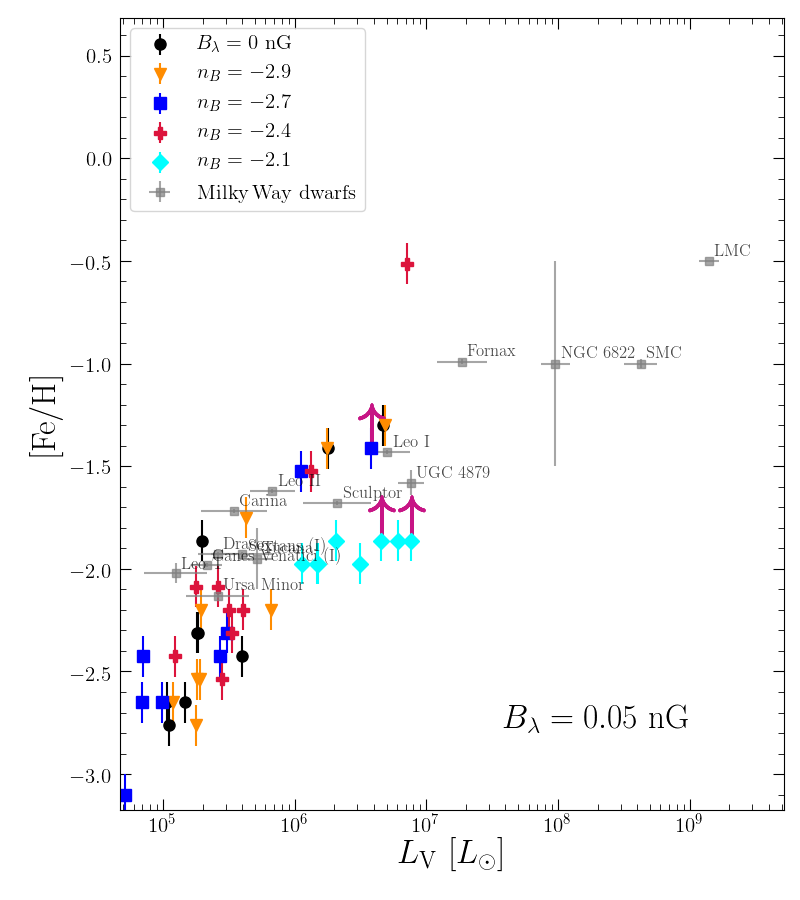}}
    {\caption{Mean metallicity vs. stellar V-band Luminosity. Each object corresponds to one model galaxy for which [Fe/H] is computed as the median of the galaxy stellar metallicity distribution function. Milky Way satellites are identified with black squares
    (see Sect.~\ref{scaling_relations} for references).
    The purple arrows indicate models that did not reach $z=0$ and are expected to have a slightly higher metallicity.}
    \label{fig:LvvsFeBl}}
    \ffigbox
    {\includegraphics[width=.5\textwidth]{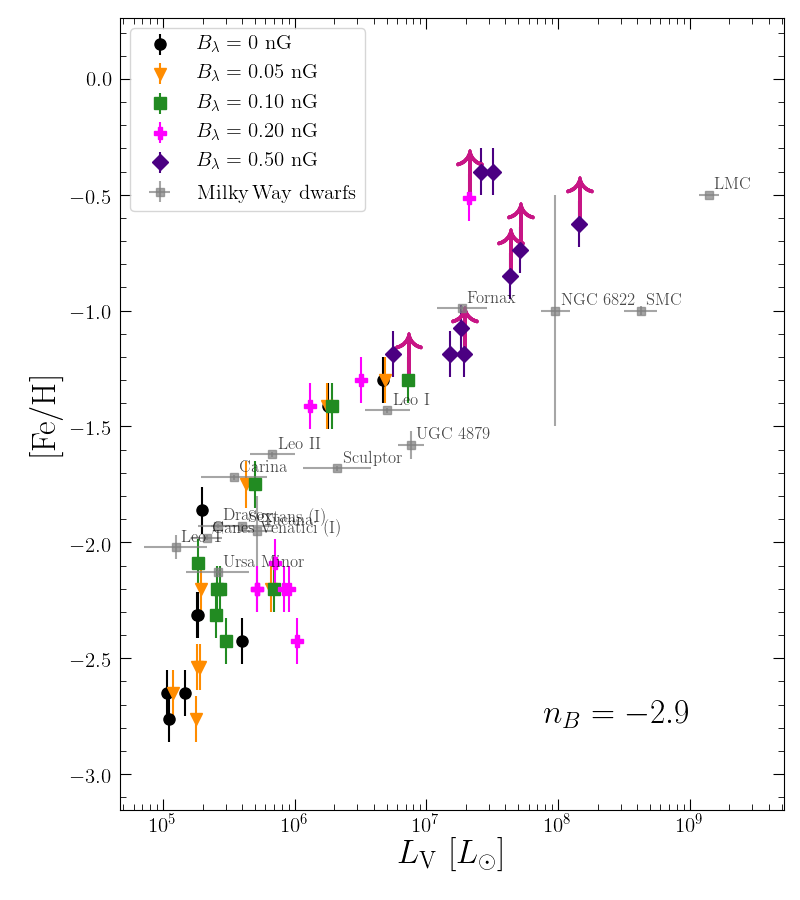}}
    {\caption{Mean metallicity vs. stellar V-band Luminosity. Same figure as Fig.~\ref{fig:LvvsFeBl} but for models with constant $n_B$ and varying 
    magnetic field strengths $B_\lambda$.}
    \label{fig:LvvsFenB}}
    \end{floatrow}
\end{figure*}{}

\begin{figure*}[t]
    \centering
    \begin{floatrow}{}
    \ffigbox
    {\includegraphics[width=.5\textwidth]{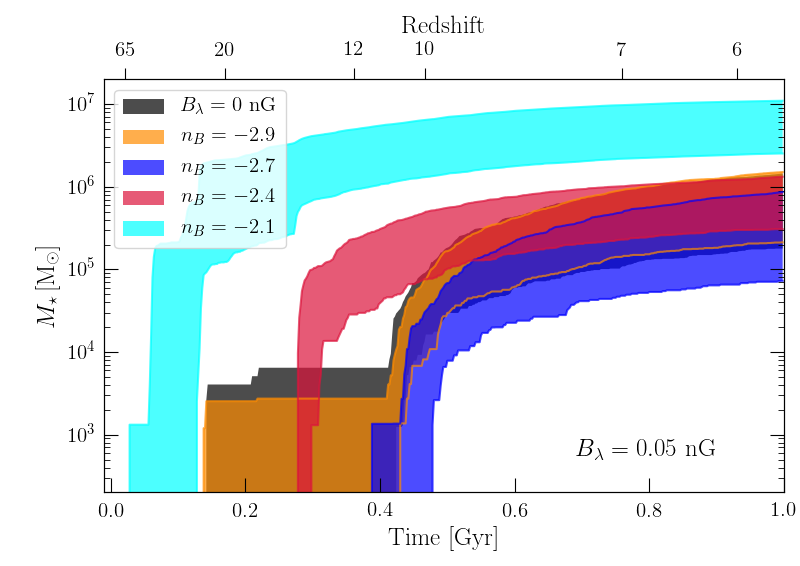}}
    {\caption
    {The cumulative stellar mass in the first Gyr for different models with varying 
    magnetic indices $n_B=-2.9$ to $-2.1$.
    The amplitude of the magnetic field is kept constant at $B_\lambda=0.05\,\textrm{nG}$. Shaded area covers all nine halos simulated in various models.}
    \label{fig:redshiftBl}}
    \ffigbox
    {\includegraphics[width=.5\textwidth]{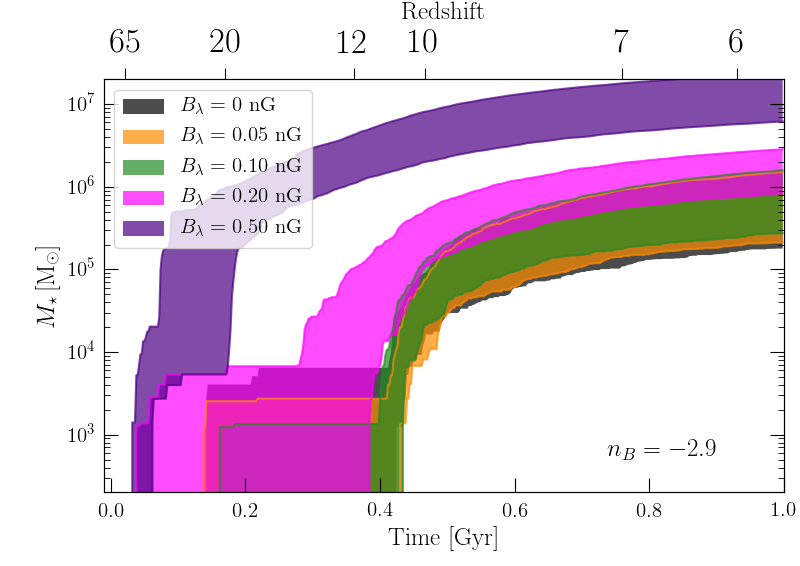}}
    {\caption{
    The cumulative stellar mass in the first Gyr. Same figure as Fig.~\ref{fig:redshiftBl} but for a 
    constant magnetic index $n_B=-2.9$ and a variety of magnetic amplitudes,
    from $B_\lambda=0.05$ to $B_\lambda=0.5\,\textrm{nG}$.}
    \label{fig:redshiftnB}}
    \end{floatrow}
\end{figure*}


\begin{figure*}[!b]
    \centering
    \begin{floatrow}{}
    \ffigbox
    {\includegraphics[width=.5\textwidth]{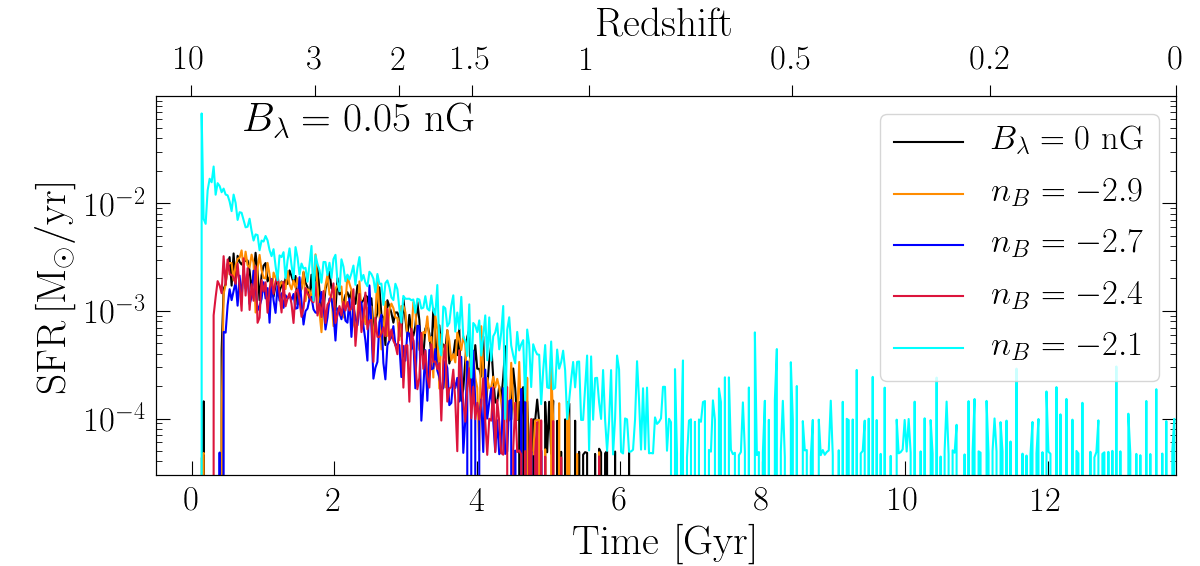}}
    {\caption
    {The star formation rate of halo \texttt{h070} for different models with varying 
    magnetic indices $n_B=-2.9$ to $-2.1$.
    The amplitude of the magnetic field is kept constant at $B_\lambda=0.05\,\textrm{nG}$.}
    \label{fig:sfrBl}}
    \ffigbox
    {\includegraphics[width=.5\textwidth]{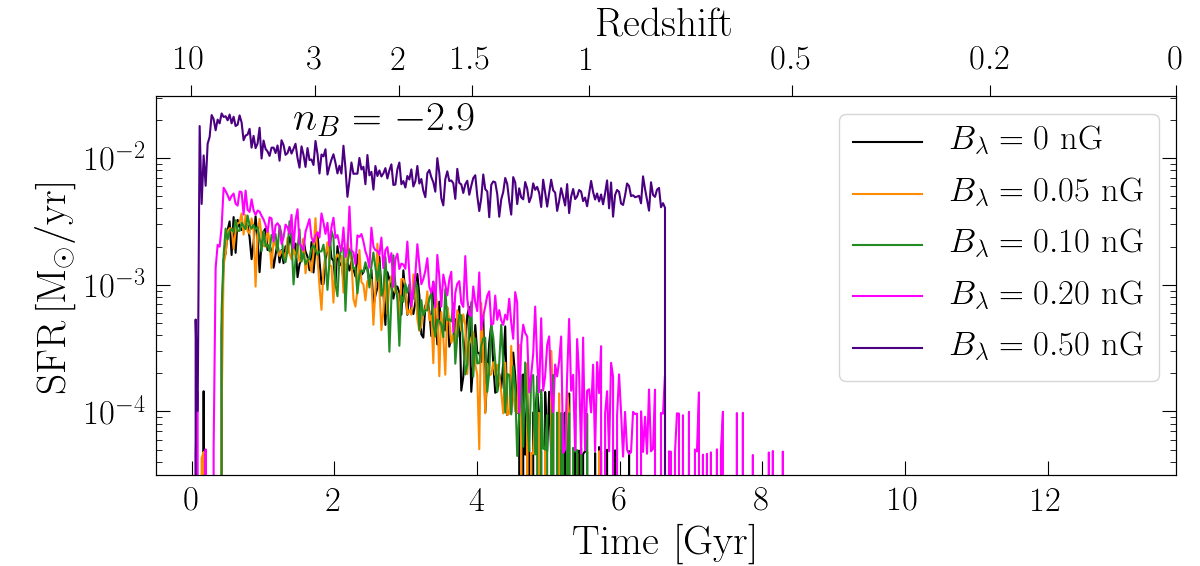}}
    {\caption{
    The star formation rate. Same figure as Fig.~\ref{fig:sfrBl} but for a 
    constant magnetic index $n_B=-2.9$ and a variety of magnetic amplitudes,
    from $B_\lambda=0.05$ to $B_\lambda=0.5\,\textrm{nG}$.}
    \label{fig:sfrnB}}
    \end{floatrow}
\end{figure*}


\begin{figure*}[t]
    \centering
    \begin{floatrow}
    \ffigbox
    {\includegraphics[width=.5\textwidth]{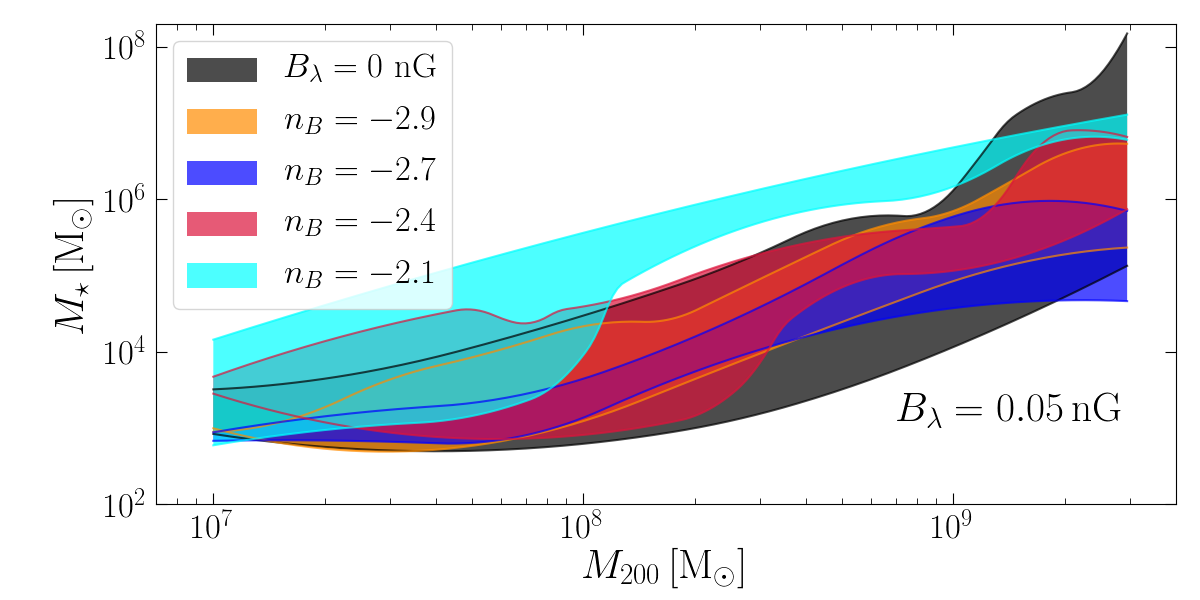}}
    {\caption{
    The stellar mass vs. halo mass relation for different models with a variety of magnetic amplitude
    $B_\lambda$ and a constant slope index $n_B=-2.9$.}
    \label{fig:LvvsMhBl}}
    \ffigbox
    {\includegraphics[width=.5\textwidth]{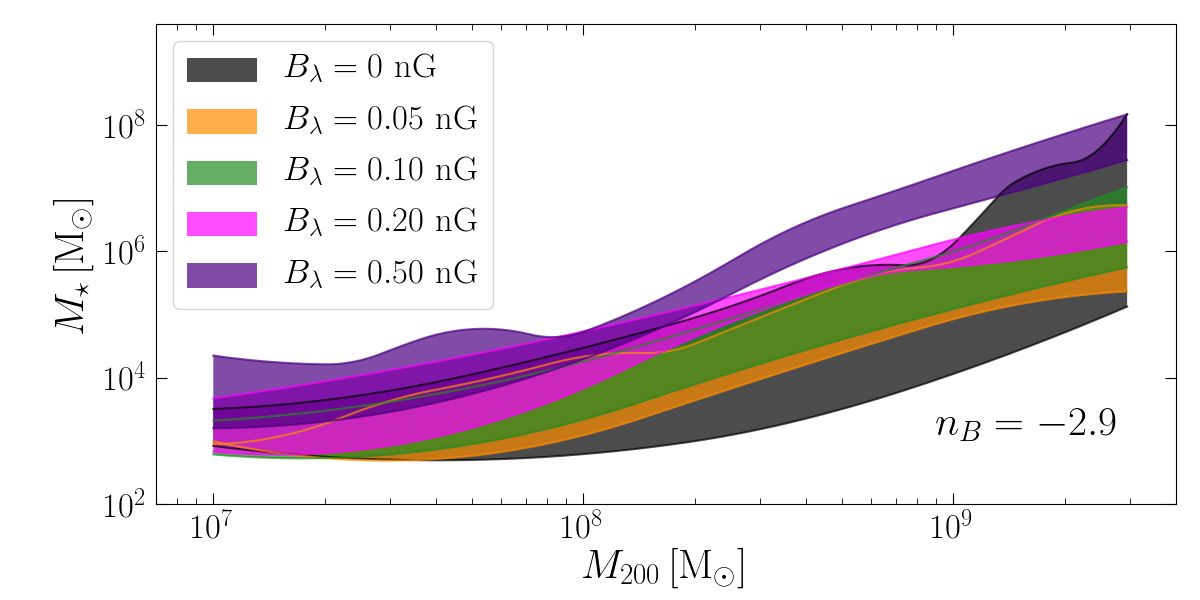}}
    {\caption{The stellar mass vs. halo mass relation. Same figure as Fig.~\ref{fig:LvvsMhBl} but for models with constant  $B_{\lambda}=0.05\,\rm{nG}$ and different slope indexes $n_B$.}
    \label{fig:LvvsMhnB}}
    \end{floatrow}
\end{figure*}

From each zoom-in simulation, we first extracted the halo corresponding to the reference halo in the unperturbed model. Then, based on the positions and particle IDs, we extracted the analogue of the same halo from the simulations in the perturbed models. 
For each extracted galaxy, we computed the following quantities defined inside one virial
radius $R_{200}$\footnote{While multiple dark sub-haloes are usually found withing $R_{200}$ at z=0, the stellar component
remain very compact ($<1\,\rm{kpc}$). We checked that extracting quantities in a smaller region limited to the size of stellar component does not impact our results.}: 
the V-band total stellar luminosity $L_{\rm{V}}$, the stellar mass $M_{\star}$, the virial mass $M_{200}$, the stellar LOS velocity dispersion $\sigma_{\rm{LOS}}$,
and the mode of the stellar metallicity distribution function [Fe/H].
The detail of the procedures used to obtain those quantities is strictly similar to the ones used in \citet{revaz2018}. 
We reported those properties in Tab.~\ref{table:phys_prop}.

Figure~\ref{fig:time_evolov} displays the time evolution of halo \texttt{h070} for five
different magnetically perturbed models
(\texttt{B0.50n2.9},  \texttt{B0.10n2.9}, \texttt{B0.05n2.1}, \texttt{B0.05n2.9}), 
from $z=11.5$ down to $z=0$. They are compared to the unperturbed case.
Table~\ref{table:phys_prop} together with Fig.~\ref{fig:time_evolov} reveal the effect of PMFs which can be
seen as a sequence along the position and amplitude of the bump in the power spectrum.

In strongly perturbed models with high amplitude $B_\lambda$, or high spectral index $n_B$, i.e.,
\texttt{B0.50n2.9}, \texttt{B0.20n2.9}, \texttt{B0.05n2.1}, the bump of the power spectrum peaks at mass 
ranges comparable to the mass scale of dwarf galaxies ($10^7$ to $10^9\,\rm{M}_{\odot}$).
Those halos start clumping, accreting gas and forming stars earlier. 
A deep potential well is quickly generated, that prevents the gas reservoir to be evaporated due to the UV-background.
Contrary to the unperturbed case where the star formation history is quenched during the epoch of reionization,
in the presence of PMFs, the star formation is extended over longer periods, up to Gyrs in the extreme cases, see Fig.~\ref{fig:sfrBl} and Fig.~\ref{fig:sfrnB}, leading to more massive and brighter galaxies at $z=0$ with a younger stellar population.
Not only the luminosity, but also the dynamics of the systems are affected,
e.g., the central LOS velocity dispersion is remarkably larger than its unperturbed analogue, see Fig.~\ref{fig:LvvsSigmaBl} and Fig.~\ref{fig:LvvsSigmanB}.
In extreme cases like \texttt{B0.50n2.9} for more massive haloes 
\texttt{h050}, \texttt{h070}, \texttt{h061}, \texttt{h141} the star formation is so intense
that it leads to a crash of the code. However, before the crash as we will see further, those
models are already too bright and too metallic to be compatible with observed dwarfs.

In weaker PMF models, namely \texttt{B0.05n2.9}, \texttt{B0.05n2.7}, as the magnetically-induced bump of the power spectrum is in a mass range much below the mass scale of dwarf galaxies ($\lesssim 10^6\,\rm{M}_{\odot}$) and the amplitude of the bump is not significant, there are no noticeable changes in their built-up history and therefore in their properties. 
The small differences we see in Tab.~\ref{table:phys_prop} are easily explained by stochasticity. For example, the star formation history is slightly expanded. Indeed, those faint systems are very sensitive to any perturbations. 

With a bump at a mass of about $10^7\,\rm{M}_{\odot}$, models \texttt{B0.05n2.4} and \texttt{B0.10n2.9} are intermediate cases.
While their luminosity and metallicity are weakly affected, a noticeable difference appears in the 
number of subhalos orbiting each dwarf at $z=0$, as seen in Fig.~\ref{fig:time_evolov}.
The increased number of those dark halos could be an excellent marker of the existence of weak PMFs. Unfortunately, nowadays, they are out of reach from observations,
as their effect on the observed properties of dwarfs is negligible and they are not massive enough
to be detected through gravitational lensing with current detection limits of about $10^9\,\rm{M}_{\odot}$ \citep{vegetti2010}.
New forthcoming facilities like the Square Kilometer Array (SKA) will allow the detection of dark substructures with masses as low as $10^6-10^7\,\rm{M}_{\odot}$
through their perturbation on gavitational arcs \citep{mckean2015}. 

\subsubsection{Zoom-in simulations: scaling relations}\label{scaling_relations}
Figure~\ref{fig:LvvsSigmaBl} and \ref{fig:LvvsSigmanB} show the galaxy V-band luminosity $L_{\rm{V}}$, as a function of the LOS velocity dispersion $\sigma_{LOS}$,
while Fig.~\ref{fig:LvvsFeBl} and \ref{fig:LvvsFenB} show their metallicity (traced by [Fe/H]) vs. $L_{\rm{V}}$. 
As comparison to the list of galaxies provided in the compilation of Local Group galaxies by \cite{mcconnachie2012}, the sample we used essentially restricted to the satellites brighter than $10^5\,\rm{L}_{\odot}$. For the [Fe/H] vs. $L_{\rm{V}}$
relation, we used only galaxies that benefit from medium resolution spectroscopy with metallicity derived either from spectral synthesis or Calcium triplet (CaT) calibration.
In \citet{revaz2018}, we showed that dwarfs emerging from a classical $\Lambda$CDM Universe (i.e., unperturbed model in this work) nicely follow those relations
over four orders of magnitude from $10^5$ to $10^9\,\rm{L}_{\odot}$. This allows us now to study the impact of PMFs.

The weakly or intermediate perturbed models (\texttt{B0.05n2.9}, \texttt{B0.05n2.7}, \texttt{B0.05n2.4}
and \texttt{B0.10n2.9}) still stay on the scaling relations.
It is worth noting that model \texttt{B0.05n2.7} suffers from a clear decrease in the luminosity (blue squares).
In this model, due to the presence of the bump in the power spectrum at a mass of $\sim 10^6\,\rm{M}_{\odot}$,
many small subhalos are formed. Contrary to more perturbed models they are not massive enough to 
form stars. This decrease in the luminosity is however not sufficient to rule out this model.
The dwarfs emerging from  intermediate case \texttt{B0.05n2.4} with their velocity dispersion being increased, fall slightly off the relation. This increase is due to the presence of a larger number of small satellite halos that dynamically heat the stellar component.

The picture becomes however much more different with stronger PMFs. 
All dwarfs simulated in models \texttt{B0.05n2.1} and \texttt{B0.50n2.9},
exhibit star formation histories longer than $2\,\rm{Gyr}$, Fig.~\ref{fig:sfrBl} and Fig.~\ref{fig:sfrnB}.
They subsequently become brighter but also denser, with strongly increased velocity dispersion.
In the extreme case of model \texttt{B0.50n2.9}, while galaxies are still following the 
$L_{\rm{V}}$ vs. $\sigma_{LOS}$ relation, they produce such a strong quantity of metals that
they lie above observed metallicity-luminosity relation, Fig.~\ref{fig:LvvsFenB}. 
\subsubsection{Luminosity vs. halo mass}\label{sec:LvvsMh}
As strongly magnetically perturbed models lead to an increase in the star formation rate, we expect the number of stars formed for a given halo mass to be modified.
While not directly observed, the stellar mass to halo mass 
relation is important in particular to predict the number of visible satellites around the Milky Way (See Sec.~\ref{sec:lg_satellites}).

To compute this relation, we extracted all halos found in the refined region
of zoom-in simulations, at $z=0$. We kept only halos containing at least ten stars and 
polluted by less than five percent of the boundary particles, i.e., particles coming from a region with a lower resolution. For each PMF model, we can then plot the stellar mass content of each halo as a function 
of its total mass, taken as $M_{200}$. The results is shown in Fig.~\ref{fig:LvvsMhBl} and \ref{fig:LvvsMhnB}.
We computed for every model an area delimited by the mean plus/minus one standard deviation of
the distribution of galaxy stellar mass withing each mass bins.
In both figures, the grey area corresponds to the unperturbed model. For the latter, to increase
the statistics, we used all dwarfs obtained from the sample of \cite{revaz2018}.

Models \texttt{B0.05n2.9}, \texttt{B0.10n2.9}, \texttt{B0.05n2.7}, and \texttt{B0.05n2.4} display
a stellar mass vs. halo mass relation which lie within the bounded area of the unperturbed case.
Model \texttt{B0.20n2.9} globally lies slightly above.
However, the two strongly perturbed models, \texttt{B0.50n2.9} and \texttt{B0.05n2.1} are clearly
above. Both produce a much larger quantity of stars, up to one dex, for a given halo mass.
\subsection{Cosmic star formation density and the reionization history of the Universe}\label{sec:reionization}

One striking impact of a strong PMF is to speed-up the formation of dark halos at the dwarf galaxy scale, where the matter
power spectrum is amplified (bumps in Fig.~\ref{fig:powerspectrumBl} and \ref{fig:powerspectrumnB}). 
This induces an earlier onset of star formation.
This effect is qualitatively seen in Fig.~\ref{fig:time_evolov} where at $z=11.5$, a dwarf is clearly
formed in models \texttt{B0.50n2.9}, \texttt{B0.05n2.1}, and \texttt{B0.05n2.4} while the gas has not  finished to collapse in the other ones yet.

To estimate this effect quantitatively, Fig.~\ref{fig:redshiftBl} and \ref{fig:redshiftnB} 
show the cumulative mass of stars formed during the first Gyr. For each model the upper (resp. the lower) edge of the coloured area corresponds to the cumulative 
mass of the dwarf which forms the larger (resp. smaller) quantity of stars.
For models \texttt{B0.50n2.9} and \texttt{B0.20n2.9} the onset of star formation starts before $z=50$
while for model \texttt{B0.05n2.1} it starts before $z=25$, contrary to the unperturbed model
in which all dwarfs form stars after $z=25$. In addition, the mean star formation rate is much stronger
for models \texttt{B0.50n2.9} and \texttt{B0.05n2.1} than the unperturbed model.

With such an early onset of star formation as well as enhancement of the star formation rate, these models will impact the reionization history of the Universe, i.e., the way the globally neutral inter galactic medium (IGM) becomes slowly ionized after being impacted by energetic UV photons generated by young and massive stars. 
The transition from a neutral IGM to an ionized one is now well constrained by different 
observations \citep[see][for a review]{fan2006b}:
the Ly-$\alpha$ absorption by the IGM (the Gunn-Peterson effect) of high redshift quasars 
\citep{fan2006a,schroeder2013,davies2018a,davies2018b,greig2019,banados2018,durovcikova2020}
or gamma-ray bursts \citep{totani2006},
the prevalence, spectral properties or clustering of Ly-$\alpha$ emitting galaxies \citep{ouchi2010,ono2012,schenker2014,tilvi2014},
and by cosmic microwave background (CMB) polarization through Thomson scattering 
\citep{spergel2003,planck2016}. 
In the following, we describe the method we used to estimate the evolution of the hydrogen 
ionized fraction from our simulations.
\subsubsection{Evolution of the hydrogen ionized fraction}
\begin{figure*}[t]
    \centering
    \begin{floatrow}{}
    \ffigbox
    {\includegraphics[width=0.5\textwidth]{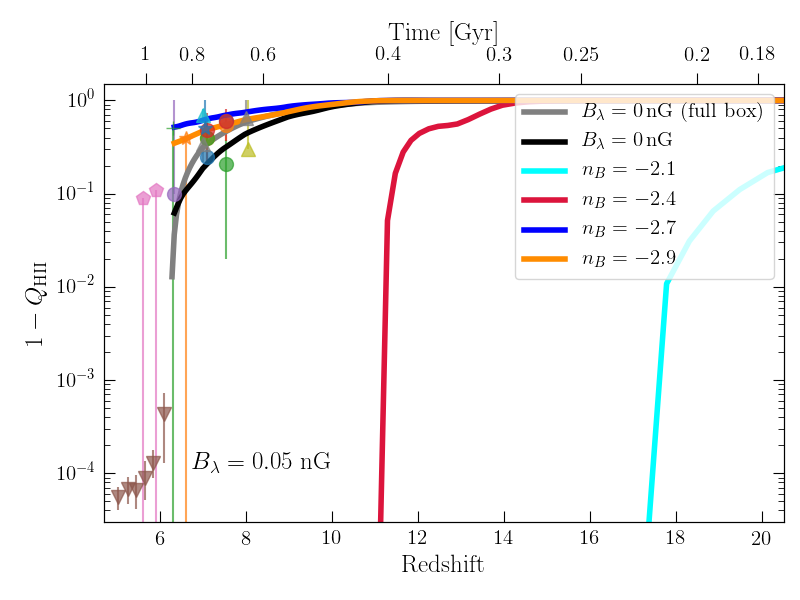}}
    {\caption
    {
    Time evolution of the neutral hydrogen fraction ($1-Q_{\rm{HII}}$) for 
    models with different slope indexes $n_B$. 
    The reference simulation (full box) is indicated by the grey curve. The model predictions are compared to observational constraints of the neutral hydrogen fraction from  Ly-$\alpha$ absorption of quasars \citep{fan2006a} in brown downward triangles, \citep{mcgreer2015} in pink pentagons, \citep{schroeder2013} in purple, \citep{davies2018a, davies2018b} in red, \citep{2017MNRAS.466.4239G,greig2019} in green, \citep{banados2018} in orange, and \citep{durovcikova2020} in blue circles, or gamma-ray bursts \citep{totani2006} in green cross and by Ly-$\alpha$ emitting galaxies \citep{ouchi2010} in orange star, \citep{ono2012} in cyan, \citep{schenker2014} in gray, and \citep{tilvi2014} in yellow triangles.
    }
    \label{fig:QHIIBl}}
    \ffigbox
    {\includegraphics[width=0.5\textwidth]{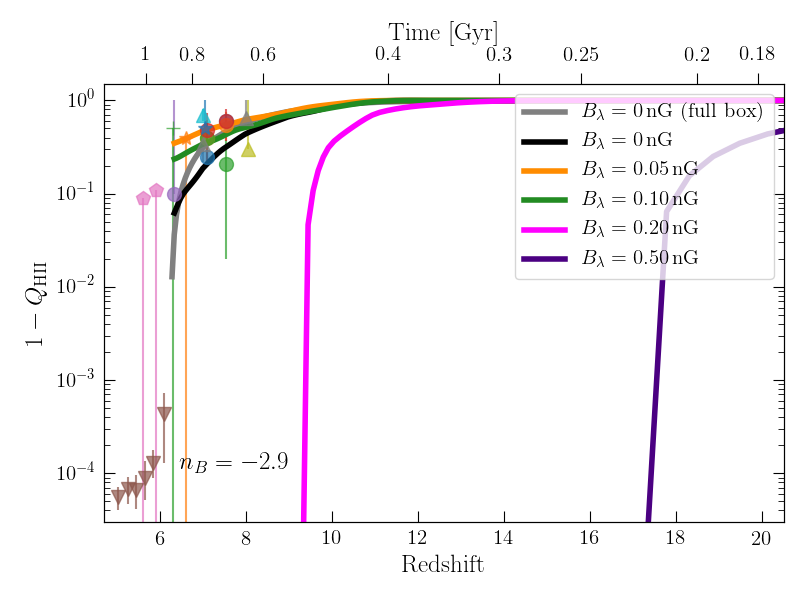}}
    {\caption{Time evolution of the neutral hydrogen fraction. Same figure as Fig.~\ref{fig:QHIIBl} but for models with varying 
    magnetic field strengths $B_{\lambda} = 0.05$ to $0.50\,\rm{nG}$.}
    \label{fig:QHIInB}}
    \end{floatrow}
\end{figure*}
The time evolution of the ionized fraction of hydrogen  $Q_{\rm{HII}}$ may be estimated using 
a simple differential equation \citep{madau1999,springel2003,kuhlen2012,stoychev2019}:
\begin{equation}
  \dot{Q}_{\rm{HII}} = \frac{\dot{n}_{\rm{ion}}}{\bar{n}_{\rm{H}}} - \frac{Q_{\rm{HII}}}{t_{\rm{rec}}}.
  \label{eq:QHII}
\end{equation}
The first term on the right hand side describes the ionization source, through the number of ionizing photons
while the second one, the sink term, is due to radiative cooling, leading to hydrogen recombination.
Further,
$\dot{n}_{\rm{ion}}$ is the global production rate of ionizing photons per unit volume, 
produced by young and massive stars. It is directly proportional to the cosmic star formation density 
$\rho_{\rm{SFR}}$:
\begin{equation}
  \dot{n}_{\rm{ion}} = \rho_{\rm{SFR}} \, f_{\rm{esc}} \, \chi_{\rm{ion}},
  \label{eq:ndot}
\end{equation}
where $f_{\rm{esc}}$ represents the fraction of ionizing photons that escape star forming regions
and $\chi_{\rm{ion}}$ the ionizing photon production efficiency for a typical stellar population
per unit time and per unit star formation rate. 
Moreover, $\bar{n}_{\rm{H}}$ is the mean comoving hydrogen number density defined through the baryonic fraction $\Omega_{\rm{b}}$, critical density $\rho_{\rm{c}}$, 
hydrogen mass fraction $X$ and hydrogen atom mass $m_{\rm{H}}$: 
\begin{equation}
  \bar{n}_{\rm{H}} = \Omega_{\rm{b}} X \rho_{\rm{c}}/m_{\rm{H}}.
\end{equation}
The source term is due to the radiative recombination of protons with electrons described by the rate of
change of the ionized hydrogen \emph{proper} density $n_{\rm{HII}}$:
\begin{equation}
  \frac{d}{d t} n_{\rm{HII}} = - n_{\rm{e}}\, n_{\rm{p}}\, \alpha_{\rm{HII}}(T),
\end{equation}
where $\alpha_{\rm{HII}}(T)$ is the temperature dependent recombination rate coefficient \citep[see][for tabulated values]{ferland1992}.
Defining $X$ and $Y$ to be the mass fraction of hydrogen and helium respectively, and $\eta$ the helium ionization degree (0, 1 or 2), we can
replace the proper number density of electron $n_{\rm{e}}$ by $n_{\rm{H}}  \left( Q_{\rm{HII}}  + \eta \frac{Y}{4\,X}\right)$ 
as well as the proper number density of proton $n_{\rm{p}}$ by $n_{\rm{H}}\,Q_{\rm{HII}} $.
The recombination time $t_{\rm{rec}}$ of Eq.~\ref{eq:QHII} is then:
\begin{equation}
  t_{\rm{rec}} = \frac{1}{ C_{\rm{HII}}\, \left(  Q_{\rm{HII}} + \eta \frac{Y}{4 X}  \right)  \bar{n}_{\rm{H}} \left( 1+z  \right)^3 \alpha_{\rm{HII}}(T) }
\end{equation}
where in the latter equation, we replaced the proper hydrogen density by its comoving expression.
An additional clumping factor $C_{\rm{HII}}$ is added to account for dense regions, self-shielded against ionizing photons that do not contribute to the recombination rate.

For each of our PMF models, we estimated the  cosmic star formation density using
the following procedure.
First, using the \texttt{Rockstar} halo finder, we extracted dark halos from our DMO simulations, 
at a redshift of $6.3$ \footnote{The choice of this redshift is dictated by a calibration
simulation that will be introduced later on.
It is sufficient to constrain our models. }.
This provides us with a complete dark halo sample covering $(3.4\,\textrm{Mpc}/h)^3$
at a redshift near the end of the Epoch of Reionization (EoR).
In a second step, using our zoom-in simulations, we extracted at the same redshift, 
all progenitors of all halos found at $z=0$ in the refined region \footnote{
Halos at $z=0$ are all halos found in the refined region containing 
at least ten stellar particles, and being polluted by less than five percent of boundary particles, i.e., particles coming from a region with a lower resolution.}.
This provides us with a sample of halos populated by galaxies for which we accurately
know the star formation history.
In a last step, we attributed to each dark halo (from the dark matter complete sample), 
a corresponding galaxy, matching their virial mass. This provides us with an estimate of the 
star formation in the full box and subsequently with the cosmic star formation density.
The main difficulty of the method is to correct for the incompleteness of the galaxy samples
extracted from the zoom-in simulations. Indeed, these samples are lacking the most
luminous objects. This does not limit our analysis as i) more massive and luminous systems 
are rare and do not dominate the ionzing photons production and ii) including them will lead to an
slightly earlier reionization, worsening the situation.

To circumvent this difficulty, we used a simulation run with the same code and parameters (including the full baryonic physics), which used the same initial perturbation field covering the same cosmological volume, but with a homogeneous resolution corresponding to the one of the refined region in the zoom-in simulations. 
This simulation allowed us to derive a reliable star formation history up to a resdhift of 6.3, where it stopped, due to computational expenses. 
We then calibrated the star formation density derived from our unperturbed zoom-in simulation samples to the full
box with a homogeneous resolution.
We found a correction factor of 3 to be sufficient to recover the complete star formation history up to $z=6.3$. We then applied this factor to the star formation density of all others models.

The evolution of the neutral hydrogen fraction $1-Q_{\rm{HII}}$ is then obtained 
from Eq.~\ref{eq:QHII}.
Setting $X=0.76$, $Y=0.24$, $\eta=1$ \citep{faucher2008}, $C_{\rm{HII}}=3$ \citep{kaurov2015},
$\chi_{\rm{ion}} =  10^{53}\,\rm{photons\,(\rm{M}_{\odot}/yr)^{-1}\,s^{-1}}$ \citep{stoychev2019}
and using a fiducial IGM temperature of $2\cdot 10^4\,\rm{K}$ \citep{hui2003}, 
data are nicely fitted by our unperturbed full box model, if the
escape fraction is set to $12\%$. Keeping this same parameters for all models, the evolution 
of the neutral hydrogen fraction $1-Q_{\rm{HII}}$ is displayed in Fig.~\ref{fig:QHIIBl} and \ref{fig:QHIInB}.

Due to a strongly enhanced star formation rates, the subsequently large amount of ionizing photons
produced by models  \texttt{B0.05n2.1}, \texttt{B0.05n2.4}, \texttt{B0.20n2.9} and \texttt{B0.50n2.9}
lead to a total reionization of the universe 
at $z < 9$. This is in total disagreement with observational constraints suggesting
a reionization starting around $z=10$ and ending at about $z=6$ \citep{2007MNRAS.382..325B, 1997ApJ...490..571L, 2000AJ....120.1167F, 1999ApJ...522L...9H}.
The other models are in much better agreement. 
We must emphasize here that our goal was not to precisely reproduce the observations. This
requires a much more precise approach, by, for example self-consistently computing the radiative transfer
of photons through the ISM and IGM. 
Our goal was rather to demonstrate that models with important magnetic field perturbation 
are far off the constraints by a comfortable margin as it is the case here.

In our approach, the growth of structures as well as the star formation
histories of dwarfs are self-consitently followed. Our results corroborate results from \citep{pandey2015}, who
estimated the Universe reionization using a semi-analytical estimation of the dark matter 
collapsed fraction, directly sensitive to matter power spectrum.
They found that models with $B_\lambda > 0.358,\,0.120,\,0.059\,\rm{nG}$ with
respectively $n_B = -2.95,\,-2.9,\,-2.85$ are ruled out from existing constraints.
Our approach also rules out models with 
$B_\lambda \geq 0.05\,\rm{nG}$ and $n_B \geq -2.4$.

\begin{figure*}[t]
    \centering
    \begin{floatrow}
    \ffigbox
    {\includegraphics[width=0.5\textwidth]{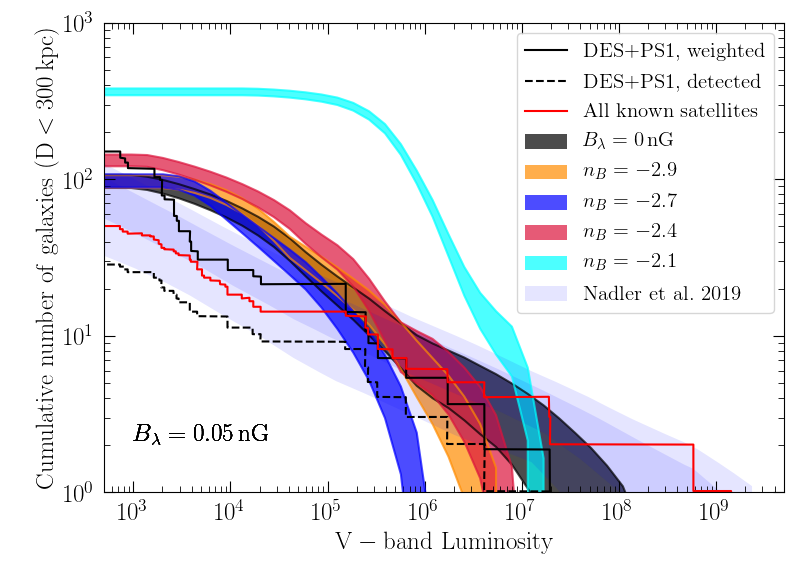}}
    {\caption{Predicted cumulative number of satellites 
    brighter than a given luminosity, within $300\,\rm{kpc}$ around the Milky Way, for 
    models with different slope indexes $n_B$. 
    For each model, the shaded area corresponds to the mean of thousand realizations of dark matter halos plus or minus one standard deviation.
    Our predictions are compared to the known satellites shown by the red curve. 
    The dashed black line shows detected satellites in Dark Energy Survey (DES) and 
    Pan-STARRS1 (PS1) as presented in \citet{drlicawagner2019}.
    The solid black line shows the same data but volume corrected, assuming 
    satellites are distributed isotropically.
    The bluish dashed regions show predictions from cosmological simulations combined with semi-analytical
    prescriptions \citep{nadler2019}
    }
    \label{fig:NCumBl}}
    \ffigbox{    \includegraphics[width=0.5\textwidth]{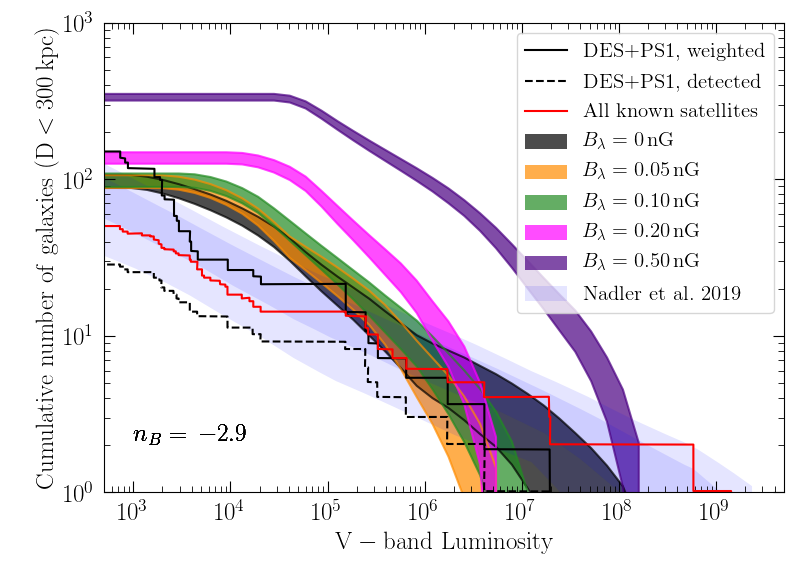}}
    {\caption{Predicted cumulative number of satellites 
    brighter than a given luminosity, within $300\,\rm{kpc}$ around the Milky Way. Same figure as Fig.~\ref{fig:NCumBl} but for models with varying 
    magnetic field strengths  $B_\lambda = 0.05$ to $0.50\,\textrm{nG}$.}
    \label{fig:NcumnB18}}
    \end{floatrow}
\end{figure*}
\subsubsection{The number of satellites in the Local Group}\label{sec:lg_satellites}

In Section~\ref{sec:halo_mass_fct} we demonstrated how PMFs may lead to the formation of
a larger number of small mass dark halos. This increase may potentially impact the number of observed
luminous satellites around the Milky Way. In this section, 
we aim at estimating  the number of expected observed dwarf galaxies brighter than a given 
luminosity inside $300\,\rm{kpc}$ and compare it with observations.
For this purpose, we combined the halo mass functions obtained in Sec.~\ref{sec:halo_mass_fct} 
with the luminosity-halo mass relations of the zoom-in hydro-dynamical simulations of Sec.~\ref{sec:LvvsMh}.

Assuming the halo mass function in the unperturbed model follows a power law (Eq.~\ref{eq:dNdM}), 
we can obtain an analytical relation of the cumulative abundance of halos $N(>M)$
by multiplying Eq.~\ref{eq:dNdM} by the magnetically induced perturbation (Eq.~\ref{eq:dNdMdNdM})
and integrate over the halo mass:

\begin{eqnarray}
     N(>M) & = &-\frac{a}{b+1}\, M_h^{b+1} + \nonumber \\ 
           & &  C \left[ 1-  \operatorname{erf}\left(\frac{\log_{10}\left( M \right)  -\left(b+1\right)s^2\ln\left(10\right)+u}{\sqrt{2}\,s}\right) \right],
\nonumber \\
          \label{eq:NM}
\end{eqnarray}
Here, $\operatorname{erf}$ is the standard error function and the constant $C$ is given by:
\begin{eqnarray}
    C & = & \frac{1}{\sqrt{2}} \sqrt{{\pi}}a  \left(c-1\right)s   \ln(10)\cdot \nonumber\\
    & & \exp{ \left(  \frac{(b+1)\ln(10) \left( (b+1)s^2\ln(10)+2u \right) }{2}  \right)}. 
\end{eqnarray}
With this notation, the unperturbed case corresponds to $C=0$.

To reproduce a realistic halo mass function of the Local Group, which includes the 
perturbative effects of both Milky Way and Andromeda galaxy, we used the cumulative number of dark matter halos inside $300\,\rm{kpc}$ predicted by the \texttt{APOSTLE} simulations \citep{sawala2017}.

The corresponding mass function is obtained by a power law with a slope $b = -1.915$ taken
by averaging the slopes of the four different radius bins given in  Tab.~2 of ~\citet{sawala2017}.
The amplitude $a=1.86\cdot 10^9$ guarantees the existence of  800 dark halos with masses larger than $10^7\,\rm{M}_{\odot}$ inside $300\,\rm{kpc}$.
This unperturbed cumulative halo distribution is then perturbed using Eq.~\ref{eq:NM}. Note that for a perturbed halo
distribution, the total number of halos with masses larger than $10^7\,\rm{M}_{\odot}$ increases (up to about 4300 for model \texttt{B0.05n2.1}), owning to the bump of the power spectrum that moves masses from smaller to larger scales.
Inverting numerically Eq.~\ref{eq:NM}, for each of our models, we randomly generated thousand realizations of dark matter halos using a Monte Carlo approach. Relying on the halo mass vs. luminosity relations (Fig.~\ref{fig:LvvsMhBl} and \ref{fig:LvvsMhnB}), we then assigned to each halo a stellar luminosity randomly chosen in the corresponding range showed by the shaded area.
Having obtained a luminosity for each halo, we computed the cumulative number of satellites 
brighter than a given luminosity.
Results are shown in Fig.~\ref{fig:NCumBl} and \ref{fig:NcumnB18}.

Except for models \texttt{B0.05n2.1}, \texttt{B0.50n2.9} and  \texttt{B0.20n2.9},
all models predict twice too many satellites for a luminosity larger than  $10^4\,\rm{L}_{\odot}$ where observed satellites show an intriguing dearth.
On the contrary, models \texttt{B0.05n2.1}, \texttt{B0.50n2.9} and  \texttt{B0.20n2.9}
are clearly above the observations, predicting the existence of respectively 16, 12 and 4 times more satellites brighter than $10^5\,\rm{L}_{\odot}$, compared to what is actually observed. 
This overabundance results from the combined effect of a larger number of dark halos expected in 
the mass $10^7-10^8\,\rm{M}_{\odot}$ (Fig.~\ref{fig:dNdMa} and \ref{fig:dNdMb}) and
slightly brighter galaxies populating a given halo mass (Fig.~\ref{fig:LvvsMhBl} and \ref{fig:LvvsMhnB}).

We finally note that models with magnetic fields seems to under-predict the number of dwarfs
brighter than $10^6\,\rm{L}_{\odot}$. This is only the result of our incomplete sample
which lacks bright dwarfs. Indeed, the dark area corresponding to the unperturbed case has been
obtained using the total sample of \citet{revaz2018} which includes dwarfs up to $5\cdot 10^8\,\rm{L}_{\odot}$. 
In this case, an excellent match is obtained with observed dwarfs.

The flattening of the curve below $10^3\,\rm{L}_{\odot}$ is the result of
the UV-background heating that evaporates gas in the smallest halos
and prevents any star formation onset as well as due to our resolution limit.

\section{Conclusion}\label{conclusions} 


We study the impact of primordial magnetic fields (PMFs) on the formation and evolution of dwarf galaxies through the modification of the $\Lambda$CDM matter power spectrum at the recombination era.
Depending on the strength and the spectral index of the magnetic field, each PMF model affects the matter power spectrum in a different mass range. We examine a variety of PMF models by either changing the amplitude ($B_\lambda=0.05, 0.10, 0.20, 0.50\,\textrm{nG}$) or the slope ($n_B=-2.9, -2.7, -2.4, -2.1$) of the magnetic power spectrum, keeping the other parameters constant. 

We first run a set of DMO simulations covering a $(3.4\,\textrm{Mpc}/h)^3$ box with a resolution of $2\times512^3$ particles.
In a second part, we re-simulate a set of nine halos extracted from the same volume, 
from redshift $z=200$ to $z=0$, using a zoom-in technique and
including a full treatment of baryons with the stellar mass resolution of $1024\,\rm{M}_{\odot}$ $h^{-1}$.
Our sample of halos in the unperturbed case, give birth to dwarf spheroidals with seven of them having a quenched star formation history and two, an extended one.
Our results are summarized as follows:

1. To quantify the contribution of PMFs in the total matter power spectrum, we compute the halo mass function for all our DMO simulations. 
The ratio of perturbed to unperturbed mass function is well fitted by a simple Gaussian function.
We show that increasing the magnetic amplitude $B_\lambda$, or the slope $n_B$, increases the number of halos around the maximum of the Gaussian function, up to a factor of 7 in the most extreme cases.

2. We extract the observable properties of each galaxy at redshift $z=0$, including, the LOS velocity dispersion $\sigma_{LOS}$, the peak metallicity, and the total V-band luminosity $L_{\rm{V}}$, 
to compare them with well observed scaling relations, such as the luminosity vs. velocity dispersion and metallicity vs. luminosity.
Strongly perturbed models, with a high amplitude ($B_\lambda=0.5\,\rm{nG}$, $n_B=-2.9$) or a steep spectral index ($n_B=-2.1$, $B_\lambda=0.05\,\rm{nG}$) have more power in the mass range
of dwarf galaxies, $10^7$ to $10^9\,\rm{M}_{\odot}$.
Consequently, in these models, galaxies form more stars and experience an extended star formation history, leading to brighter 
and more metal rich systems, incompatible with the observed Local Group scaling relations.
On the contrary, the observed properties of the weakly perturbed models are
not sensitively modified.

3. We show that strong magnetic models speed up the structure formation with an impact on the reionization of the Universe. We estimate the fraction of hydrogen ionized in the first Gyr of the Universe history and demonstrate that earlier onset of star formation as well as the higher rate of forming stars in these models, induces a large amount of ionizing photons, enough to reionize the universe at redshifts higher than $z=9$, incompatible with the observational constraints for the EoR.

4. By combining the abundance of dark matter halos obtained from DMO simulations with the luminosity-halo mass function, we derive the number of luminous satellites expected around the Milky Way and show that with high magnetic amplitude or spectral index not only the number of small dark matter halos in mass ranges between $10^7$ and $10^8\,\rm{M}_{\odot}$ is raised, but also the stellar mass content for a given halo mass is increased, resulting in an overabundance of satellites brighter than $10^5\,\rm{L}_{\odot}$, contradicting with observations in the Local Group. 

We conclude that galaxies simulated in weakly perturbed models resemble all physical properties of their counterparts in the unperturbed model. However, stronger models such as, $B_\lambda=0.05\,\rm{nG}$, with $n_B=-2.4$ or $n_B=-2.1$, and $B_\lambda=0.20\,\rm{nG}$, or $B_\lambda=0.5\,\rm{nG}$, with $n_B=-2.9$ may be ruled out due to the aforementioned reasoning. Our results are consistent with cosmological observables, namely, CMB observations, weak gravitational lensing, Lyman$-\alpha$ data, etc.,  that constrain the amplitude and the spectral index of the magnetic field power spectrum \citep{2004PhRvD..69f3006C, 2004PhRvD..70d3011L, 2010PhRvD..82h3005K, 2012ApJ...748...27P, 2012PhRvL.108w1301T, 2012PhRvD..86d3510S, 2013ApJ...762...15P}. We show that even smaller perturbed models can have a detectable impact on the structure formation. 

In this study we considered the contribution of PMFs in the $\Lambda$CDM matter power spectrum in mass scales comparable to dwarf galaxies. 
Regarding the possible direct impact of magnetic fields intrinsic to galaxies, in the proto-stellar scales there are different procedures involved in boosting the star formation rate namely by the angular momentum loss due to the magnetic braking \citep{MouschoviasPaleologou1980,10.1046/j.1365-8711.2000.03215.x}, or in contrary suppressing the star formation due to the magnetic pressure \citep{BurkhartEtAl2009, MolinaEtAl2012, MoczEtAl2017, 2020arXiv200211502S}. 
Though, the net effect is still an open question (See \citet{2019FrASS...6....7K} for a recent review).
It is certainly worth studying whether the modification to the star formation history due to the above procedures dominate over the impact of PMFs during the formation of the first structures. To answer that including the full magneto-hydrodynamics (MHD) treatment in our simulations is required. However
it is too computationally expensive at this moment. We let this improvement for future studies.


\begin{acknowledgements}
We would like to thank Loïc Hausammann, Mladen Ivkovic, and Florian Cabot for very useful discussions.    
We gratefully acknowledges financial support by Swiss government scholarship FCS. 
J.S. acknowledges the funding from the
European Unions Horizon 2020 research and
innovation program under the Marie Sk{\l}odowska-Curie Grant
No.\ 665667 and
the support by the Swiss National Science Foundation under Grant No.\ 185863.
K.E.K. acknowledges financial support by the Spanish Science Ministry grant PGC2018-094626-B-C22.
We acknowledges the support by the International Space Science Institute (ISSI), Bern, Switzerland, for supporting and funding the international team “First stars in dwarf galaxies”. 
This work was supported by the Swiss
Federal Institute of Technology in Lausanne (EPFL) through the use of the
facilities of its Scientific IT and Application Support Center (SCITAS). The simulations presented here were run on the Deneb clusters. The data
reduction and galaxy maps have been performed using the parallelized Python
\texttt{pNbody} package (\url{http://lastro.epfl.ch/projects/pNbody/}).
\end{acknowledgements}

\bibliographystyle{aa}
\bibliography{ms}
\clearpage
\onecolumn
\begin{appendix}{}
\section{Physical Properties of model galaxies}

\begin{longtable}[c]{l c c c c c c c c r}
\captionsetup{width=.85\textwidth}
\caption{ Physical properties of all model galaxies.
For each galaxy we computed 
the V-band total stellar luminosity $L_{\rm{V}}$, 
the stellar mass $M_\star$,
the virial mass $M_{200}$,
the virial radius $R_{200}$,
the stellar line-of-sight velocity dispersion $\sigma_{\rm{LOS}}$,
and the mode of the stellar metallicity distribution function [Fe/H], all defined inside one virial 
radius. In the first row, different PMF models are shown at a variety of magnetic spectral indices between $-2.9$ to $-2.1$ for constant $B_\lambda = 0.05$, and in the second row for a constant spectral index at $n_B\ =\ -2.9$ and magnetic field strengths between $B_\lambda\ =\ 0.05 - 0.50\ \textrm{nG}$. }\label{table:phys_prop}

\\
\hline\hline

model & $B_\lambda$ & $n_B$ & $L_{\rm{V}}$ & $M_\star$ & $M_{200}$ & $R_{200}$ & $\sigma_\textrm{{LOS}}$ & [Fe/H] & Redshift \footnote{halos which did not reach $z = 0$ crashed at redshift specified here. The reason some simulations crashed at $z > 0$ is due the increase in the mass content of more massive halos in strong models which consequently leads to extremely massive galaxies with a very early onset and high rate of star formation.} / Time\\
ID & $[rm{nG}]$ & - & $[10^6\rm{L}_{\odot}]$ & $[10^6\rm{M}_{\odot}]$ & $[10^9\rm{M}_{\odot}]$ & kpc & $[\rm{kms}^{-1}]$ & dex & - / [Gyr]\\
\hline
\endfirsthead
\caption{continued.}\\
\hline\hline

model & $B_\lambda$ & $n_B$ & $L_{\rm{V}}$ & $M_\star$ & $M_{200}$ & $R_{200}$ & $\sigma_{\rm{LOS}}$ & [Fe/H] & Redshift / Time\\
ID & $[\rm{nG}]$ & - & $[10^6\rm{L}_{\odot}]$ & $[10^6\rm{M}_{\odot}]$ & $[10^9\rm{M}_{\odot}]$ & kpc & $[\rm{kms}^{-1}]$ & dex & - / [Gyr]\\
\hline
\endhead
\hline
\endfoot
 &  &  &  &  & \texttt{h050} &  &  &  &  \\
 \hline
 \texttt{B0.00n0.0} & 0.00 & 0.0 & 4.66 & 10.01 & 2.66 & 33.14 & 10.52 & -1.30 & 0.00 / 13.8 \\
\hline
\texttt{B0.05n2.9} & 0.05 & -2.9 & 4.90 & 10.47 & 2.67 & 33.18 & 10.80 & -1.30 & 0.00 / 13.8 \\
\texttt{B0.05n2.7} &  & -2.7 & 3.84 & 7.60 & 2.63 & 32.98 & 11.30 & -1.41 & 0.06 / 13.0 \\
\texttt{B0.05n2.4} &  & -2.4 & 7.10 & 12.29 & 2.46 & 32.25 & 12.09 & -0.51 & 0.00 / 13.8 \\
\texttt{B0.05n2.1} &  & -2.1 & - & - & - & - & - & - & 24.6 / 0.13  \\
\hline
\texttt{B0.10n2.9} & 0.10 & -2.9 & 7.33 & 13.76 & 2.62 & 32.94 & 11.61 & -1.30 & 0.04 / 13.3 \\
\texttt{B0.20n2.9} & 0.20 &  & 20.99 & 31.93 & 2.51 & 32.49 & 14.35 & -0.51 & 0.11 / 12.3 \\
\texttt{B0.50n2.9} & 0.50 &  & 143.40 & 147.52 & 3.51 & 36.33 & 25.83 & -0.62 & 0.62 / 7.8 \\
\hline
 &  &  &  &  & \texttt{h070} &  &  &  &  \\
 \hline
 \texttt{B0.00n0.0} & 0.00 & 0.0 & 1.78 & 5.13 & 1.80 & 29.08 & 10.20 & -1.41 & 0.00 / 13.8 \\
\hline
\texttt{B0.05n2.9} & 0.05 & -2.9 & 1.76 & 5.05 & 1.78 & 28.99 & 10.11 & -1.41 & 0.00 / 13.8 \\
\texttt{B0.05n2.7} &  & -2.7 & 1.11 & 3.13 & 1.81 & 29.13 & 10.57 & -1.52 & 0.00 / 13.8 \\
\texttt{B0.05n2.4} &  & -2.4 & 1.34 & 3.84 & 1.93 & 29.74 & 13.46 & -1.52 & 0.00 / 13.8 \\
\texttt{B0.05n2.1} &  & -2.1 & 6.13 & 17.16 & 1.61 & 28.02 & 15.61 & -1.86 & 0.00 / 13.8 \\
\hline
\texttt{B0.10n2.9} & 0.10 & -2.9 & 1.93 & 5.52 & 1.79 & 29.00 & 10.44 & -1.41 & 0.00 / 13.8 \\
\texttt{B0.20n2.9} & 0.20 &  & 3.21 & 9.34 & 1.84 & 29.32 & 12.00 & -1.30 & 0.00 / 13.8 \\
\texttt{B0.50n2.9} & 0.50 &  & 51.22 & 53.89 & 2.38 & 31.90 & 18.28 & -0.74 & 0.83 / 6.6 \\
\hline
 &  &  &  &  & \texttt{h061} &  &  &  &  \\
\hline
\texttt{B0.00n0.0} & 0.00 & 0.0 & 0.20 & 0.50 & 1.95 & 29.85 & 9.72 & -1.86 & 0.00 / 13.8 \\
\hline
\texttt{B0.05n2.9} & 0.05 & -2.9 & 0.18 & 0.43 & 1.96 & 29.91 & 10.10 & -2.76 & 0.00 / 13.8 \\
\texttt{B0.05n2.7} &  & -2.7 & 0.04 & 0.09 & 1.98 & 30.00 & 10.16 & -3.89 & 0.00 / 13.8 \\
\texttt{B0.05n2.4} &  & -2.4 & 0.28 & 0.75 & 1.79 & 29.02 & 12.00 & -2.54 & 0.00 / 13.8 \\
\texttt{B0.05n2.1} &  & -2.1 & 7.68 & 13.37 & 1.80 & 29.05 & 14.81 & -1.86 & 0.75 / 7.0 \\
\hline
\texttt{B0.10n2.9} & 0.10 & -2.9 & 0.30 & 0.74 & 1.94 & 29.82 & 10.25 & -2.42 & 0.00 / 13.8 \\
\texttt{B0.20n2.9} & 0.20 &  & 1.05 & 2.75 & 2.02 & 30.23 & 13.10 & -2.42 & 0.00 / 13.8 \\
\texttt{B0.50n2.9} & 0.50 &  & 43.60 & 48.94 & 2.59 & 32.82 & 17.66 & -0.85 & 0.69 / 7.4 \\
\hline
 &  &  &  &  & \texttt{h141} &  &  &  &  \\
\hline
\texttt{B0.00n0.0} & 0.00 & 0.0 & 0.18 & 0.49 & 0.78 & 21.98 & 8.22 & -2.31 & 0.00 / 13.8 \\
\hline
\texttt{B0.05n2.9} & 0.05 & -2.9 & 0.20 & 0.52 & 0.78 & 21.97 & 8.36 & -2.20 & 0.00 / 13.8 \\
\texttt{B0.05n2.7} &  & -2.7 & 0.07 & 0.18 & 0.78 & 22.05 & 8.73 & -2.42 & 0.00 / 13.8 \\
\texttt{B0.05n2.4} &  & -2.4 & 0.26 & 0.72 & 0.87 & 22.78 & 13.17 & -2.09 & 0.00 / 13.8 \\
\texttt{B0.05n2.1} &  & -2.1 & 4.52 & 8.39 & 0.84 & 22.53 & 12.94 & -1.86 & 0.67 / 7.5 \\
\hline
\texttt{B0.10n2.9} & 0.10 & -2.9 & 0.27 & 0.71 & 0.81 & 22.31 & 8.33 & -2.20 & 0.00 / 13.8 \\
\texttt{B0.20n2.9} & 0.20 &  & 0.71 & 1.95 & 0.79 & 22.10 & 10.43 & -2.09 & 0.00 / 13.8 \\
\texttt{B0.50n2.9} & 0.50 &  & 19.26 & 20.80 & 1.08 & 24.49 & 15.42 & -1.19 & 1.13 / 5.4 \\
\hline
 &  &  &  &  & \texttt{h111} &  &  &  &  \\
\hline
\texttt{B0.00n0.0} & 0.00 & 0.0 & 0.15 & 0.36 & 1.06 & 24.40 & 10.71 & -2.65 & 0.00 / 13.8 \\
\hline
\texttt{B0.05n2.9} & 0.05 & -2.9 & 0.18 & 0.45 & 1.05 & 24.26 & 9.13 & -2.54 & 0.00 / 13.8 \\
\texttt{B0.05n2.7} &  & -2.7 & 0.10 & 0.23 & 1.07 & 24.43 & 11.99 & -2.65 & 0.00 / 13.8 \\
\texttt{B0.05n2.4} &  & -2.4 & 0.34 & 0.91 & 0.93 & 23.29 & 13.26 & -2.31 & 0.00 / 13.8 \\
\texttt{B0.05n2.1} &  & -2.1 & 3.13 & 8.98 & 0.88 & 22.92 & 16.41 & -1.98 & 0.00 / 13.8 \\
\hline
\texttt{B0.10n2.9} & 0.10 & -2.9 & 0.26 & 0.64 & 1.06 & 24.38 & 10.15 & -2.20 & 0.00 / 13.8 \\
\texttt{B0.20n2.9} & 0.20 &  & 0.83 & 2.21 & 1.19 & 25.32 & 11.94 & -2.20 & 0.00 / 13.8 \\
\texttt{B0.50n2.9} & 0.50 &  & 32.33 & 65.88 & 1.95 & 29.86 & 18.98 & -0.40 & 0.00 / 13.8 \\
\hline
 &  &  &  &  & \texttt{h122} &  &  &  &  \\
\hline
\texttt{B0.00n0.0} & 0.00 & 0.0 & 0.11 & 0.26 & 0.93 & 23.37 & 9.02 & -2.65 & 0.00 / 13.8 \\
\hline
\texttt{B0.05n2.9} & 0.05 & -2.9 & 0.12 & 0.30 & 0.96 & 23.61 & 9.22 & -2.65 & 0.00 / 13.8 \\
\texttt{B0.05n2.7} &  & -2.7 & 0.05 & 0.12 & 0.89 & 22.98 & 10.53 & -3.10 & 0.00 / 13.8 \\
\texttt{B0.05n2.4} &  & -2.4 & 0.12 & 0.32 & 0.86 & 22.77 & 13.81 & -2.42 & 0.00 / 13.8 \\
\texttt{B0.05n2.1} &  & -2.1 & 1.48 & 4.13 & 0.84 & 22.56 & 15.05 & -1.98 & 0.00 / 13.8 \\
\hline
\texttt{B0.10n2.9} & 0.10 & -2.9 & 0.18 & 0.46 & 0.91 & 23.17 & 8.77 & -2.09 & 0.00 / 13.8 \\
\texttt{B0.20n2.9} & 0.20 &  & 0.52 & 1.36 & 0.90 & 23.12 & 10.92 & -2.20 & 0.00 / 13.8 \\
\texttt{B0.50n2.9} & 0.50 &  & 25.93 & 53.81 & 1.54 & 27.60 & 18.33 & -0.40 & 0.00 / 13.8 \\
\hline
 &  &  &  &  & \texttt{h159} &  &  &  &  \\
\hline
\texttt{B0.00n0.0} & 0.00 & 0.0 & 0.40 & 1.03 & 0.68 & 21.02 & 9.31 & -2.42 & 0.00 / 13.8 \\
\hline
\texttt{B0.05n2.9} & 0.05 & -2.9 & 0.43 & 1.09 & 0.68 & 21.05 & 8.96 & -1.75 & 0.00 / 13.8 \\
\texttt{B0.05n2.7} &  & -2.7 & 0.27 & 0.69 & 0.67 & 20.93 & 10.23 & -2.42 & 0.00 / 13.8 \\
\texttt{B0.05n2.4} &  & -2.4 & 0.40 & 1.10 & 0.59 & 20.09 & 11.59 & -2.20 & 0.00 / 13.8 \\
\texttt{B0.05n2.1} &  & -2.1 & 2.08 & 6.06 & 0.65 & 20.73 & 15.12 & -1.86 & 0.00 / 13.8 \\
\hline
\texttt{B0.10n2.9} & 0.10 & -2.9 & 0.50 & 1.29 & 0.68 & 21.00 & 9.98 & -1.75 & 0.00 / 13.8 \\
\texttt{B0.20n2.9} & 0.20 &  & 0.91 & 2.49 & 0.67 & 20.93 & 10.49 & -2.20 & 0.00 / 13.8 \\
\texttt{B0.50n2.9} & 0.50 &  & 15.10 & 36.82 & 1.64 & 28.18 & 18.10 & -1.19 & 0.00 / 13.8 \\
\hline
 &  &  &  &  & \texttt{h168} &  &  &  &  \\
\hline
\texttt{B0.00n0.0} & 0.00 & 0.0 & 0.11 & 0.27 & 0.54 & 19.42 & 8.68 & -2.76 & 0.00 / 13.8 \\
\hline
\texttt{B0.05n2.9} & 0.05 & -2.9 & 0.66 & 1.79 & 2.88 & 34.01 & 9.60 & -2.20 & 0.00 / 13.8 \\
\texttt{B0.05n2.7} &  & -2.7 & 0.31 & 0.81 & 2.70 & 33.29 & 10.70 & -2.31 & 0.00 / 13.8 \\
\texttt{B0.05n2.4} &  & -2.4 & 0.32 & 0.88 & 1.14 & 24.93 & 11.90 & -2.20 & 0.00 / 13.8 \\
\texttt{B0.05n2.1} &  & -2.1 & 1.51 & 4.43 & 0.52 & 19.18 & 11.08 & -1.98 & 0.00 / 13.8 \\
\hline
\texttt{B0.10n2.9} & 0.10 & -2.9 & 0.70 & 1.92 & 2.68 & 33.20 & 10.15 & -2.20 & 0.00 / 13.8 \\
\texttt{B0.20n2.9} & 0.20 &  & 1.31 & 3.67 & 2.31 & 31.61 & 12.15 & -1.41 & 0.00 / 13.8 \\
\texttt{B0.50n2.9} & 0.50 &  & 18.36 & 40.09 & 1.43 & 26.94 & 18.34 & -1.08 & 0.00 / 13.8 \\
\hline
 &  &  &  &  & \texttt{h177} &  &  &  &  \\
\hline
\texttt{B0.00n0.0} & 0.00 & 0.0 & 0.18 & 0.47 & 0.52 & 19.26 & 8.35 & -2.31 & 0.00 / 13.8 \\
\hline
\texttt{B0.05n2.9} & 0.05 & -2.9 & 0.19 & 0.50 & 0.52 & 19.22 & 7.53 & -2.54 & 0.00 / 13.8 \\
\texttt{B0.05n2.7} &  & -2.7 & 0.07 & 0.17 & 0.51 & 19.07 & 8.02 & -2.65 & 0.00 / 13.8 \\
\texttt{B0.05n2.4} &  & -2.4 & 0.18 & 0.49 & 0.37 & 17.08 & 10.72 & -2.09 & 0.00 / 13.8 \\
\texttt{B0.05n2.1} &  & -2.1 & 1.14 & 3.18 & 0.68 & 21.06 & 13.11 & -1.98 & 0.00 / 13.8 \\
\hline
\texttt{B0.10n2.9} & 0.10 & -2.9 & 0.25 & 0.65 & 0.54 & 19.42 & 8.27 & -2.31 & 0.00 / 13.8 \\
\texttt{B0.20n2.9} & 0.20 &  & 0.52 & 1.40 & 0.56 & 19.67 & 8.64 & -2.20 & 0.00 / 13.8 \\
\texttt{B0.50n2.9} & 0.50 &  & 5.63 & 17.27 & 0.94 & 23.41 & 13.54 & -1.19 & 0.00 / 13.8 \\
\hline

\end{longtable}

\footnotetext{halos which did not reach $z = 0$ crashed at redshift specified here. The reason some simulations crashed at $z > 0$ is due the increase in the mass content of more massive halos in strong models which consequently leads to extremely massive galaxies with a very early onset and high rate of star formation.}
\clearpage

\end{appendix}

\end{document}